\documentclass[showpacs,twocolumn,aps,prd]{revtex4-1}
\usepackage{graphicx}
\usepackage{dcolumn}
\usepackage{bm}
\usepackage{amsmath}
\usepackage{amssymb}
\usepackage[usenames,dvipsnames]{xcolor}
\usepackage{color}
\usepackage[colorlinks,plainpages=false]{hyperref}

\begin{document}
  \title{Magnetized strangelets with anomalous magnetic moment and Coulomb interactions}
\author{Huai-Min~Chen,$^1$}
\email{chenhuaimin@wuyiu.edu.cn}

\author{Xiao-Wei~Li,$^3$}
\email{12031243@mail.sustech.edu.cn}
\author{Cheng-Jun~Xia$^{4}$}
\email{cjxia@yzu.edu.cn}
\author{Jing-Tao~Wang$^{2}$}

\author{Guang-Xiong~Peng$^{2,5,6}$}
\email{gxpeng@ucas.ac.cn}

\affiliation{%
    $^1$\mbox{School of Mechanical and Electrical Engineering, Wuyi University, Wuyishan 354300, China}\\
	$^2$\mbox{School of Nuclear Science and Technology, University of Chinese Academy of Sciences, Beijing 100049, China}\\
    $^3$\mbox{Shenzhen Institute for Quantum Science and Engineering, Southern University of Science and Technology, Shenzhen 518055, China}\\
	$^4$\mbox{Center for Gravitation and Cosmology, College of Physical Science and Technology, Yangzhou University, Yangzhou 225009, China}\\
	$^5$\mbox{Theoretical Physics Center for Science Facilities, Institute of High Energy Physics, P.O. Box 918, Beijing 100049, China}\\
	$^6$\mbox{Synergetic Innovation Center for Quantum Effects and Application, Hunan Normal University, Changsha 410081, China}  }

\begin{abstract}
We study the magnetized strangelets in the baryon density-dependent quark mass model, including the effects of both confinement and lead-order perturbation interactions.
The properties of magnetized strangelets are investigated under the the field strength $2\times 10^{17}\ \mathrm{G}$, where the anisotropy caused by the strong magnetic field is insignificant can be treated approximately as an isotropic system.
The consideration of anomalous magnetic moments in the energy spectrum naturally solves the difficulty of infrared divergence encountered in integrating the density of states.
The Coulomb interaction is accounted for a self-consistent treatment. The energy per baryon, mechanically stable radius, strangeness and electric charge of magnetized strangelets are presented, where their dependence on the field strength and parameter of confinement and perturbation are investigated.
\end{abstract}

\maketitle
\section{Introduction}
\label{intro}
Strange quark matter(SQM) is one type of matter comprised of deconfined up, down and strange quarks, which could be the ground state of QCD, rather than $^{56}\mathrm{Fe}$~\cite{Madsen_-162}. This idea is first suggested by Bodmer in 1971~\cite{Bodmer1971_PRD4-1601}, and developed by Witten in 1984~\cite{Witten1984_PRD30-272}. The small lumps of SQM whose baryon number $A\lesssim10^{7}$ is called strangelets, named by Berger and Jaffe~\cite{Berger1987_PRC35-213}, which is so small that electron can not exist in its interior, as the electron Compton's wavelength is much larger than the radius of strangelets~\cite{Madsen_-162}.

It is possible that strangelets can be produced by high-energy cosmic rays collide with each other or the core of neutron stars~\cite{Finch2006_JPG32-s251,Biswas2017_PRC95-045201,Annala2020,Xia_PhysRevD.102.023031}, strange star collisions could release strangelets as part of the energetic cosmic rays~\cite{Banerjee_JPG.25.L15,Banerjee_PhysRevL.85.1384,Madsen_PhysRevL.90.121102,Madsen_PhysRevD.71.014026,Shaulov_EPJWebC.52.04010}.
In addition, strangelets could also be found as the products of high energy heavy ion collision~\cite{Schaffner-Bielich1997_PRC55-3038,Alimena2020_JPG47-090501,Parvu2021_JCAP11-040}.

The strong magnetic field could exist in heavy-ion collisions and compact stars, e.g., pulsars and magnetars. For pulsars, the typical magnitudes of surface magnetic fields are about $10^{14}\mathrm{G}$~\cite{Duncan1992_AJL392-L9,Taylor1993_AJSS88-529,Sob'yanin2023_PRD107-L081301}, and the observed magnetic field strength on the surface of magnetars could be $10^{14}\sim10^{15}\mathrm{G}$~\cite{Kaspi2017_55-261}. Moreover, the strong magnetic fields might be generated in non-central high-energy heavy ion collisions, reaching the value of $10^{19}\mathrm{G}$~\cite{Mizher2010_PRD82-105016}.

 Due to the difficulties of perturbation calculations at low energy and high density region and the sign problem in lattice point simulations, the study of quark matter depends on various phenomenological models, e.g., the magnetized strange quark matter (MSQM) with bag model~\cite{CHAKRABARTY1994_09-3611,Felipe2008_PRC77-015807,Tomoki2020_JPCS}, the Nambu-Jona-Lasinio (NJL) model~\cite{Mizher2010_PRD82-105016,Li2016_PRD93-054005,Maruyama2019_JCP}, the mass-density-dependent model~\cite{Hou2015_CPC39-015101}, the quark quasiparticle model~\cite{Wen2013a}, the confined-isospin-density-dependent mass model~\cite{Chu2018_PLB778-447}, and the magnetized strangelets with the bag model~\cite{Felipe2012_39-045006,Ding2014_62-859}. At present, the MIT bag model has been widely used and obtained some important results. Madsen considered the finite size effects of strangelets~\cite{Madsen_PhysRevLett.70.391}. Lugones and Grunfeld studied the effects of surface tension and vector interactions on the existence of a strangelet crust in the MIT Bag model~\cite{GL2021prc103,GL2021prd104}. Recently, they have revisit the quark-mass density-dependent model and show that thermodynamic inconsistencies that have plagued the model for decades can be solved if the model is formulated in the canonical ensemble instead of the grand canonical one~\cite{GL_PhysRevD.107.043025}. Presently, we adopt a thermodynamically self-consistent mass-density-dependent model, so called baryon density-dependent quark mass model~\cite{Chen_PhysRevD.105.014011,Chen_CPC.46.055102}, to investigate the properties of magnetized strangelets at zero temperature.

As is shown in Ref.~\cite{Ferrer2010_PRC82-065802}, due to the existence of uniform magnetic field, the $\emph{O}(3)$ rotation symmetry breaks into $\emph{O}(2)$ symmetry, and therefore distinction between longitudinal- and transverse-to-the-field pressures is caused. Nevertheless, we assume a spherical symmetry of strangelets when magnetic field strength $B\lesssim2\times 10^{17}\mathrm{G}$. Using the method adopted in Ref.~\cite{Ferrer2010_PRC82-065802,Felipe2012_39-045006}, we calculate the quantity $\delta=|P_{\parallel}-P_{\perp}|/|P(B=0)|$, and use it as a measure to distinguish between isotropic and anisotropic regions~\cite{Felipe2012_39-045006}.

The electric charge of strangelets is an important property of strangelets. In the absence of electron, generally, strangelets carry slight charge. Some authors propose that positively charged strangelets would repel nuclei, while negatively charged ones would attract them, leading to catastrophic transition from normal matter into SQM~\cite{Madsen_-162}. We treat the Coulomb interaction in a thermodynamicly self-consistent manner, where the chemical potential contributed by the Coulomb energy is included.

The present paper is organized as follow. In Sec.~\ref{Therm}, we give the thermodynamic treatment of strangelets in a strong magnetic field with baryon density-dependent quark mass model, while, the contribution of Coulomb interaction is accounted for in a thermodynamically self-consistent approach. In Sec.~\ref{sec:mass}, we adopt the quark mass scaling at zero temperatures, where both the confinement interaction and perturbation interaction are considered with the equivalent mass of quarks. In Sec.~\ref{Properties}, the numerical results of the magnetized strangelets are presented, e.g., their energy per baryon, radius and charges, as well as the dependence of energy per baryon and radius on the parameters of mass scaling. Finally, a short summary is given in Sec.~\ref{Summ}.
\section{Thermodynamic treatment in a strong magnetic field}
\label{Therm}
\subsection{Thermodynamic framework}
  The thermodynamic potential density $\Omega_0$ at zero temperature is given for a free quasiparticle system due to its dependence on $\mu^*$, i.e.,
 \begin{equation}\label{omega}
   \Omega_{0}=\sum_{i}\int \Big(\sqrt{p^{2}+m_{i}^{2}}-\mu_{i}^{*}\Big)\rho_{i}(p)\mathrm{d}^{3}p,
 \end{equation}
 where index $i$ goes over the particle types of system, and $\rho_{i}$ is the density of state. In previous quark mass-density-dependent model, quark acts like a free particle with a density-dependent mass $m_{i}=m_{0,i}+m_{I}(n_{b})$, where $n_{b}=\frac{1}{3}\sum_{i}n_{i}$ and $m_{0,i}$ are baryon number density and the quark current mass with $m_{u0}=5\ \mathrm{MeV}$, $m_{d0}=10\ \mathrm{MeV}$, and $m_{s0}=100\ \mathrm{MeV}$. In terms of the multiple reflection expansion method~\cite{Madsen_PhysRevLett.70.391,Balian1970_AoP60-401},
  the function $\rho_{i}$ of strangelets can be given by
 \begin{equation}\label{rho}
   \rho_{i}=\frac{g_{i}}{(2\pi)^{3}V}\Big[V+\frac{S}{p}f_{S}\big(\frac{m_{i}}{p}
   \big)+\frac{C_{r}}{p^{2}}f_{C}\big(\frac{m_{i}}{p}\big)\Big],
 \end{equation}
 where $V$, $S$ and $C_{r}$ are volume, area and extrinsic curvature of strangelets, and $g_{i}$ is the corresponding degree of freedom. For a spherical system, we have $V=4\pi R^{3}/3$, $S=4\pi R^{2}$, $C_{r}=8\pi R$. The functions $f_{S}$ and $f_{C}$ are given by~\cite{Berger1987_PRC35-213,Madsen1994_PRD50-3328}
 \begin{eqnarray}\label{func}
   f_{S}&=&-\frac{1}{2}\mathrm{arctan}\big(\frac{m_{i}}{p}\big),\\
   f_{C}&=&\frac{1}{6}\Big[1-\frac{3p}{2m_{i}}\mathrm{arctan}\big(\frac{m_{i}}{p}
   \big)\Big].
 \end{eqnarray}

Note that the density of states in Eq.~(\ref{rho}) becomes negative and is hence unphysical at small radii and momenta. An infrared cutoff~\cite{Neergaard1999_PRD60-054011} is usually introduced to treat the unphysical contributions, which is not included here since their contributions are small and the natural truncation caused by anomalous magnetic moments has considered.

At the zero temperature, within the framework of an equivalent mass model~\cite{Peng_PhysRevC.62.025801}, the particle mass depends on baryon number density $n_b$. Given that the baryon number density is defined as $n_b \equiv \sum_i N_i/(3V)$, this implies that the particle mass becomes a function of volume and particle numbers. In such a model, if we adopt the expressions of free particle for the thermodynamic potential density and chemical potential as presented in Eq.~(\ref{omega}) to represent the actual values for the system, it would lead to thermodynamic inconsistencies~\cite{Xia2014_PRD89-105027}.

To avoid this issue, it is necessary to incorporate the corrections arising from the dependence of mass on the state variables. Based on the consideration, the analysis starts with the energy $\overline{E}(V,\{N_i\})=V\int\sqrt{p^2+m_i^2}\rho_i \text{d}^3p$, which is a function of the state variables such as volume and particle number.
At the zero temperature, the energy is given by
\begin{equation}\label{eq:free}
    \overline{E}(V,\{N_{i}\},\{m_{i}\})=\Omega_0 V + \sum_i\mu_i^* N_i.
\end{equation}
Considering mass as an intermediate variable, the differential of Eq.~(\ref{eq:free}) is given by
\begin{eqnarray}\label{eq:diff2}
\text{d}\overline{E} &= & \left[\Omega_0+V\left(\frac{\partial \Omega_0}{\partial V}\right)_{\{m_i\},\{N_i\}}\right]\text{d}V+\sum_i \mu_i^*\text{d}N_i \nonumber \\
&+ & \sum_i V\left(\frac{\partial \Omega_0}{\partial m_i}\right)_{\{N_i\}, V}\text{d}m_i.
\end{eqnarray}
The differential of $m_i$ is as follows
\begin{equation}
    \begin{aligned}
        \text{d}m_i&=\frac{\partial m_i}{\partial n_b}\left[\left(\frac{\partial n_b}{\partial V}\right)_{N_i}\text{d}V+\sum_{j}\left(\frac{\partial n_b}{\partial N_j}\right)_V\text{d}N_j\right]\\
        &=\frac{1}{3V^2}\sum_j\frac{\partial m_i}{\partial n_b}(V \text{d}N_j-N_j\text{d}V).
    \end{aligned}
\end{equation}

The fundamental thermodynamic differential relations of energy is
\begin{equation}\label{eq:diff1}
    \text{d}\overline{E}=-P\text{d}V+\sum_i \mu_i \text{d}N_i,
\end{equation}
where $P$, $V$, $\mu_i$ and $N_i$ correspond to the pressure, volume, chemical potential and particle number of the system, respectively.

For a uniform system, the energy density $E=\overline{E}/V$, and particle number density $n_i = N_i/V$.
Comparing Eq.~(\ref{eq:diff2}) with Eq.~(\ref{eq:diff1}), we can derive the following thermodynamic relationships
\begin{eqnarray}
        P &= & -\Omega_{0}-V\left(\frac{\partial \Omega_0}{\partial V}\right)_{\{m_i\},\{N_i\}}  \nonumber \label{eq:P} \\
        &+ &n_b\sum_j\frac{\partial m_j}{\partial n_b}\left(\frac{\partial \Omega_0}{\partial m_j}\right)_{\{N_i\}, V},\\
        \mu_i &=& \mu_i^* +\frac{1}{3}\sum_j\frac{\partial m_j}{\partial n_b}\left(\frac{\partial \Omega_0}{\partial m_j}\right)_{\{N_i\}, V}.\label{eq:mu}
\end{eqnarray}
Based on $E=\Omega_{0}+\sum_{i}\mu^{*}_{i}n_{i}=\Omega+\sum_{i}\mu_{i}n_{i}$, we have
\begin{eqnarray}\label{eq:E}
    \Omega=\Omega_0 -n_b\sum_j\frac{\partial m_j}{\partial n_b}\left(\frac{\partial \Omega_0}{\partial m_j}\right)_{\{N_i\}, V},
\end{eqnarray}
where the second term is under a fixed volume, i.e., $\partial \Omega_0/\partial V=\partial \Omega/\partial V$. By Eqs.~(\ref{eq:P})-(\ref{eq:E}), we obtained
\begin{eqnarray}
E &= & -P+\sum_{i}\mu_{i}n_{i}-V\left(\frac{\partial \Omega_0}{\partial V}\right)_{\{m_i\},\{N_i\}}\nonumber \\
&=& -P+\sum_{i}\mu_{i}n_{i}-V\left(\frac{\partial \Omega}{\partial V}\right)_{\{m_i\},\{N_i\}},
\end{eqnarray}
\begin{eqnarray}
P &=& -\Omega-V\left(\frac{\partial \Omega_0}{\partial V}\right)_{\{m_i\},\{N_i\}} \nonumber \\
&=& -\Omega-V\left(\frac{\partial \Omega}{\partial V}\right)_{\{m_i\},\{N_i\}}.
\end{eqnarray}
Because of the presence of the surface and curvature terms, the thermodynamic potential density needs to be derived for the volume. When $R$ is large enough, the surface and curvature terms tend to vanish, i.e., $\partial \Omega/\partial V=0$. The system transitions from strangelets to strange quark matter in such large $R$, thus $E = - P + \sum_{i} \mu_{i} n_{i}$ and $P = - \Omega$. The Euler equation at zero temperature is validated.

By Eq.~(\ref{eq:diff1}), we have
\begin{eqnarray}
P = -\left.\frac{d(VE)}{\mbox{d}V}\right|_{\{N_i\}}.
\end{eqnarray}
i.e.,
\begin{eqnarray}
P &=& -\frac{d(VE)}{\mbox{d}n_\mathrm{b}} \left.\frac{\mbox{d}n_\mathrm{b}}{dV}\right|_{\{N_i\}} \nonumber \\
& =& - \frac{N\mbox{d}(E/n_\mathrm{b})}{\mbox{d}n_\mathrm{b}} \left(-\frac{N}{V^2}\right) \nonumber \\
&=& \frac{N^2}{V^2} \frac{\mbox{d}(E/n_\mathrm{b})}{\mbox{d}n_\mathrm{b}} \nonumber \\
&=& n_\mathrm{b}^2 \frac{\mbox{d}(E/n_\mathrm{b})}{\mbox{d}n_\mathrm{b}}.
\end{eqnarray}
It can be easily seen from the relationship that the pressure must be zero at the minimum energy per baryon, which provides a simple and intuitive method for testing thermodynamic self consistency.

Based on $E(V,\{n_{i}\},\{m_{i}\})=\Omega_{0}+\sum_{i}\mu^{*}_{i}n_{i}=\Omega+\sum_{i}\mu_{i}n_{i}$, we have $\partial E/\partial \mu_{i}^{*}=\partial \Omega_{0}/\partial \mu_{i}^{*} + n_{i}=0$. The particle number densities is
\begin{eqnarray}
n_{i} = -\frac{\partial\Omega_{0}}{\partial \mu_{i}^{*}}. \label{10}
\end{eqnarray}
This is the same as the formula derived from the equivalent particle model~\cite{Xia2014_PRD89-105027}.

\subsection{Charged fermion in a uniform magnetic field}
In the non-relativistic case, the energy levels of charged particles under a uniform magnetic field were given by Landau in 1930. Transitioning from classical theory to quantum mechanics, taking into account the spin of charged particles, the Hamiltonian operator in electromagnetic fields is
\begin{equation}\label{HEM}
  \hat{H}=\frac{1}{2m}\Big(\hat{{\bf p}}-q{\bm A}\Big)^{2}-\hat{{\bm \mu}}\cdot{\bm B}+q\phi,
\end{equation}
where ${\bm B}$ is the magnetic field strength (in the natural system of units, taking into account that the vacuum permeability is taken to be 1, the magnetic field strength is equal to the magnetic induction strength ${\bm H}={\bm B}$), ${\bm A}$ is the magnetic vector potential, and $\phi$ is the electric potential. The intrinsic magnetic moment $\hat{{\bm \mu}}=\mu\hat{{\bm s}}/s$, where $\mu$ is the magnetic moment value, $\hat{{\bm s}}$ is the particle spin operator, and $s$ is the spin quantum number. Let the uniform magnetic field~${\bm B}=(0,0,B)$, then the vectorial potential can be taken as
\begin{equation}\label{Ai}
  A_{x}=-By,\quad A_{y}=A_{z}=0.
\end{equation}
The electric potential is taken to be 0 and the Hamiltonian operator becomes
\begin{equation}\label{HM}
  \hat{H}=\frac{1}{2m}\Big(\hat{p_{x}}+qBy\Big)^{2}+\frac{\hat{p}_{y}^{2}+\hat{p}_{z}^{2}}{2m}-\frac{\mu}{s}\hat{s}_{z}B.
\end{equation}
Noting that $\hat{s}_{z}$ is commutative with the Hamiltonian, this operator can be replaced by the eigenvalue $\sigma$, then the spin part of the wave function does not matter, and the Schr$\ddot{\mathrm{o}}$dinger equation is taken to be
\begin{equation}\label{SH}
  \frac{1}{2m}\Big[\Big(\hat{p_{x}}+qBy\Big)^{2}+\hat{p}_{y}^{2}+\hat{p}_{z}^{2}\Big]\psi-\frac{\mu}{s}\sigma B\psi=E\psi.
\end{equation}
Further note that the Hamiltonian is commutative with ~$\hat{p}_{x}$ and ~$\hat{p}_{z}$ and $\psi$ can be taken as $\psi=\mathrm{exp}(i(p_{x}x+p_{z}z))\chi(y)$. The ~$\chi$ satisfies
\begin{equation}\label{SC}
  \chi''+2m\Big[\Big(E+\frac{\mu\sigma B}{s}-\frac{p_{z}^{2}}{2m}\Big)-\frac{1}{2}m\omega^{2}(y-y_{0})^{2}\Big]\chi=0,
\end{equation}
where $y_{0}=-p_{x}/qB$, $\omega=|q|B/m$. Let $E'=E+\mu\sigma B/s-p_{z}^{2}/2m$, then Eq.~(\ref{SC}) is equivalent to the equation of a linear harmonic oscillator with energy $E'$ and frequency $\omega$. The energy eigenvalues of the linear harmonic oscillator are $E'=(n+1/2)\omega$, $n=0,1,2,\cdots$.

As a result, the energy level of a charged particle in a uniform magnetic field is
\begin{equation}\label{Landaulevel}
E_{n}=\Big(n+\frac{1}{2}\Big)\omega+\frac{p_{z}^{2}}{2m}-\frac{\mu\sigma B}{s}.
\end{equation}
The first term in the above equation corresponds to the motion in the transverse plane, and these energy levels $n$ are known as Landau energy levels. For charged fermions, $\mu/s=\mathrm{sgn}(q)|q|/m$, where $\mathrm{sgn}(q)$ is the sign function. Thereby, Eq.~(\ref{Landaulevel}) can be rewritten as
\begin{equation}\label{Landauenergy}
  E_{p,q}=\frac{p_{z}^{2}}{2m}+\frac{e|q|B}{2m}\Big[2n+1-2\sigma\mathrm{sgn}(q)\Big]=\frac{p^{2}}{2m}.
\end{equation}
In the case of relativity, the energy of particles
\begin{equation}\label{Renergy}
  E_{p}=\sqrt{p^{2}+m^{2}}.
\end{equation}
Taking $\eta=2\sigma\mathrm{sgn}(q)$, it can be seen from (\ref{Landauenergy}) that the transverse momentum is $p_{\perp}=\sqrt{p_{x}^{2}+p_{y}^{2}}=\sqrt{|q|B(2n+1-\eta)}$. Thus, the energy of a charged particle in a uniform magnetic field in the relativistic case should be
\begin{equation}\label{RMenergy}
  E_{p_{z},q}=\sqrt{p_{z}^{2}+e|q|B(2n+1-\eta)+m^{2}}.
\end{equation}
For multi-body systems, to determine the state in which the system is in, it is usually necessary to integrate in momentum space. Notice that the transverse momentum is discretized, and the integration in the transverse direction should become a summation. Such that $2\nu=2n+1-\eta$. The rules for substituting integrals for summation given by~\cite{Chakrabarty1996_PRD54-1306}
\begin{equation} \label{Rule-0}
  \int_{-\infty}^{\infty}\int_{-\infty}^{\infty}\mathrm{d} p_{x}\mathrm{d} p_{y}\longrightarrow 2\pi e|q_{i}|B\sum_{\nu=0}^{\infty}(2-\delta_{\nu,0}),
  \end{equation}

In addition to having an impact on particle energy levels and momentum, a strong magnetic field can also cause the pressure distribution of the system to be anisotropic. The pressure also split in $P_{\parallel}$ and $P_{\perp}$, which denote parallel and transverse to the magnetic field direction~\cite{Ferrer2010_PRC82-065802}. The expressions of $P_{\parallel}$ and $P_{\perp}$ for a
magnetized fermion system can be written as
\begin{eqnarray}
  P_{\parallel} &=& \sum_{i}\mu_{i}n_{i}-E-V\left(\frac{\partial \Omega}{\partial V}\right), \label{EPP}\\
  P_{\perp} &=& \sum_{i}\mu_{i}n_{i}-E-V\left(\frac{\partial \Omega}{\partial V}\right)+B^{2}-\mathcal{M}B. \label{EPPT}
\end{eqnarray}
where $\mathcal{M}$ is the magnetization of the system, which is given by
  \begin{equation}
    \mathcal{M}=-\frac{\partial \Omega}{\partial B}.
  \end{equation}
Details of the derivation of the parallel and transverse pressures from the energy-momentum tensor are provided in Appendix \ref{sec:app}.
It can been see that the parallel pressure $P_{\parallel}$ satisfies the Hugenholtz-Van Hove (HVH) theorem~\cite{HUGENHOLTZ1958363}, while the transverse pressure $P_{\perp}$ has extra contributions from the magnetic field, which will result in the zero-pressure point density being consistent with the density at the minimum of the energy per baryon for $P_{\parallel}$ but not for $P_{\perp}$~\cite{Chu_PhysRevD.90.063013}.

Considering a strong magnetic field, the energy density at zero temperature is given by~\cite{Ferrer2010_PRC82-065802}
  \begin{eqnarray}\label{eqenergyE}
    E&=&\Omega_{0}+\sum_{i}\mu_{i}^{*}n_{i}+\frac{B^{2}}{2}\nonumber\\
     &=&\Omega+\sum_{i}\mu_{i}n_{i}+\frac{B^{2}}{2}.
  \end{eqnarray}

Due to $\partial \Omega_0/\partial V=\partial \Omega/\partial V$, the parallel and transverse pressures in a strong magnetic field are given by
  \begin{eqnarray}
    P_{\parallel}&=&-\Omega-V\frac{\partial \Omega}{\partial V}-\frac{B^{2}}{2} \nonumber \\
    &=&-\Omega-V\frac{\partial \Omega_{0}}{\partial V}-\frac{B^{2}}{2}, \label{pressureP}\\
    P_{\perp}&=&-\Omega-V\frac{\partial \Omega}{\partial V}+\frac{B^{2}}{2}-\mathcal{M}B \nonumber \\
    &=&-\Omega-V\frac{\partial \Omega_{0}}{\partial V}+\frac{B^{2}}{2}-\mathcal{M}B. \label{pressureT}
  \end{eqnarray}

We can use a quantity $\delta$ to quantify the anisotropy of SQM, which is defined by
\begin{equation}\label{delta}
    \delta=|\frac{P_{\parallel}-P_{\perp}}{P(B=0)}|.
\end{equation}
We assume the spherical shape for strangelets and $P_{\parallel}\simeq P_{\perp}$ when $\delta\lesssim 0.1$. For SQM, it up to magnetic field values approximated of $2\times10^{17}\ \mathrm{G}$~\cite{Felipe2012_39-045006} when $\mu_{d}^{*}=\mu_{s}^{*}=\mu_{u}^{*}+\mu_{e}=400\mathrm{MeV}$.
In our calculation, we adopt that the mechanical stability condition is $P_{\parallel}=0$ for $B\leq 2\times10^{17}\ \mathrm{G}$ since the strangelets is a self bound system, which is equivalent to minimize the total free energy at fixed $N_i$ as usually adopted in the literature~\cite{Madsen_PhysRevD.50.3328, Lugones2011_PRD84-085003}.

\subsection{Anomalous magnetic moment}
Strictly speaking, the energy formula under the relativistic case should be given by the Dirac equation, and, in addition to the intrinsic magnetic moments, the anomalous magnetic moments of charged particles should be taken into account. In 1950, Johnson and Lippman first considered the anomalous moments in the Dirac equation~\cite{Johnson1950_PR77-702}. However, the formula they gave was not covariant. Bjorken and Drell provided a formula for covariation
\begin{equation}\label{AMM}
  \Big[\gamma^{\mu}\Big(\partial_{\mu}+i|q|A_{\mu}\Big)-\frac{i\mu}{2}F_{\mu\nu}\gamma^
  {\mu}\gamma^{\nu}+m
  \Big]\psi=0.
\end{equation}
Then, the energy of charged particles containing anomalous magnetic moments can be obtained
\begin{equation}\label{EnergyAMM}
  E=\sqrt{p_{z}^{2}+\Big(\sqrt{e|q|B(2n+1-\eta)+m^{2}}-\eta|Q|B\Big)^{2}}.
\end{equation}
It should be noted that $(n=\nu,\eta=1)$ and $(n=\nu-1,\eta=-1)$ no longer equivalent due to the existence of anomalous magnetic moment $Q$, then the integral replacement rule Eq.~(\ref{Rule-0})  becomes
\begin{equation} \label{Rule-1}
  \int_{-\infty}^{\infty}\int_{-\infty}^{\infty}\mathrm{d} p_{x}\mathrm{d} p_{y}\longrightarrow 2\pi e|q_{i}|B\sum_{\eta=\pm 1}\sum_{n=0}^{\infty}.
  \end{equation}

Considering that the components of the quark momentum must be real and that the Fermi momentum $\nu_{q}=\sqrt{\mu_{q}^{*2}-m_{q}^{2}}$ is an upper limit to the quark momentum at zero temperature, there must be
\begin{equation}\label{Realp}
\mu_{q}^{*2}-m_{q}^{2}-p_{\perp}^{2}\geq 0.
\end{equation}
By Eq.~(\ref{Realp}), we have
\begin{equation}\label{Neq}
  n\leq \frac{\mu_{q}^{*2}-m_{q}^{2}}{2e|q_{q}|B}+\frac{\eta-1}{2}.
\end{equation}
Thus, the upper limit for the sum of Landau energy levels in Eq.~(\ref{Rule-1}) is
\begin{equation}\label{Nmax}
  n_{\mathrm{max}}=\mathrm{int}\Big[\frac{\mu_{q}^{*2}-m_{q}^{2}}{2e|q_{q}|B}\Big]+\frac
  {\eta-1}{2},
\end{equation}
where $\mathrm{int}[x]$ is an integer function that means taking the integer part of argument $x$.

The anomalous magnetic moments of electrons and quarks are given
\begin{equation}\label{AMMQ}
\begin{split}
  Q_{e}=0.0016\mu_{B},\quad &Q_{u}=1.85\mu_{N},\\
  Q_{d}=-0.97\mu_{N},\quad  &Q_{s}=-0.58\mu_{N},
\end{split}
\end{equation}
where $\mu_{B}\simeq5.79\times10^{-15}\ \mathrm{MeV/G}$, $\mu_{N}\simeq3.15\times10^{-18}\ \mathrm{MeV/G}$.
We can see that these anomalous magnetic moments are very small, and unless in extremely strong magnetic field environments, the contribution of anomalous magnetic moments can be completely ignored. Taking $Q=0$, the Eq.~(\ref{EnergyAMM}) returns to Eq.~(\ref{RMenergy}). The contribution of anomalous magnetic moments is not significant, but they play a crucial role in the study of the properties of strangelets under strong magnetic fields.

If the anomalous magnetic moment is not considered, the integration of the density of states under a magnetic field will encounter difficulties in infrared divergence due to the presence of surface and curvature terms. When $n=0$ and $\eta=1$, there are two infrared divergences $\mathrm{ln}(p_{\perp, i})$ and $\mathrm{ln}(p_{\perp, i})-1/p_{\perp, i}$ for density of states under magnetic field, caused by the second and third term of Eq.~(\ref{rho}), so called surface term and curvature term, respectively.

To address this issue, we introduce an infrared cutoff for $p_z$. The anomalous magnetic moment provides a natural cutoff $(2m-|Q|B)|Q|B$ ~\cite{Felipe2008_PRC77-015807}, meaning that when $n=0$ and $\eta=1$, the lower limit of the integral for $p_z$ becomes $(2m-|Q|B)|Q|B$. The natural cutoff arises due to the mass-shell condition
\begin{equation}\label{eq:mass_shell}
    p^2=E^2-m^2\geq 0.
\end{equation}
When $n=0$ and $\eta=1$, Eq.~(\ref{EnergyAMM}) gives us
\begin{equation}\label{eq:amm}
    E=\sqrt{p_z^2+(m-|Q|B)^2}.
\end{equation}
Substituting Eq.~(\ref{eq:amm}) into Eq.~(\ref{eq:mass_shell}), we obtain
\begin{equation}
    p_z^2+m^2-2m|Q|B+|Q|^2B^2-m^2\geq 0,
\end{equation}
i.e.,
\begin{equation}
  p_{z}\geq\sqrt{(2m-|Q|B)|Q|B}.
\end{equation}
In essence, under the condition $n=0$ and $\eta=1$, the minimum allowed value of $p_z$
is determined by the ``natural cutoff" derived from the interplay between the particle's effective mass $m$ and the external magnetic field $B$. This ``natural cutoff" circumvents infrared divergences arising from surface and curvature terms.

Considering the anisotropy caused by a strong magnetic field and the anomalous magnetic moment of quarks, the Eq.~(\ref{omega}) becomes
  \begin{equation} \label{Omega00}
    \Omega_{0}=\sum_{i}\sum_{\eta=\pm 1}\sum_{n=0}^{n_{max}}2\pi e|q_{i}|B\int_{-p_{i}^{(n,\eta)}}^{p_{i}^{(n,\eta)}}(E_{p,i}-\mu_{i}^{*})\rho_{i}\mathrm{d}p_{z},
  \end{equation}
  where $p_{i}^{(n,\eta)}=\sqrt{p_{F,i}^{2}-p_{\perp}^{2}}$, $p_{F,i}=\sqrt{\mu_{i}^{*2}-m_{i}^{2}}$ is the Fermi momenta of quark with flavor $i$. Due to the very small value of the anomalous magnetic moment, the contribution of the anomalous magnetic moment to the energy levels is negligible. Taking $Q=0$ in Eq.~(\ref{EnergyAMM}), the energy spectrum $E_{p,i}$ of the quark with flavor $i$ is
   \begin{equation}   \label{eqEp}
    E_{p,i}=\sqrt{p_{z}^{2}+p_{\perp}^{2}+m_{i}^2},
  \end{equation}
  \begin{equation}    \label{eqpt}
    p_{\perp}=m_{i}\sqrt{\frac{B}{B_{i}^{c}}(2n+1-\eta)},\qquad B_{i}^{c}=\frac{m_{i}^{2}}{e|q_{i}|},
  \end{equation}
  i.e.,
  \begin{equation}    \label{eqpts}
    p_{\perp}=\sqrt{e|q_{i}|B(2n+1-\eta)}.
  \end{equation}
  Here, $m_{i}$ are the quark masses, $n$ indexes the Landau level, $\eta=2\mathrm{sgn}(q_{i})s=\pm 1$ correspond to the orientations of the particle magnetic moment parallel or antiparallel to the magnetic field. The sign function $\mathrm{sgn}(x)$ equals to $1$ with a positive argument and to $-1$ with a negative argument, $s=\pm\frac{1}{2}$ are spin projections onto the magnetic field direction. $B_{i}^{c}$ are the critical magnetic field~\cite{Felipe2008_PRC77-015807}, $q_{i}$ denote the quark electric charges(e.g., $q_{u}=2/3$, $q_{d}=q_{s}=-1/3$), $e$ represents the elementary
  charge. Note that the degeneracy breaking down caused by magnetic field, the statistical weight $g_{i}=3$ for color of quarks.\par

Consequently, the number densities of quarks are given by
  \begin{eqnarray} \label{number}
    n_{i}&=& -\frac{\partial \Omega_{0}}{\partial \mu_{i}^{*}}\nonumber\\
         &=&\sum_{\eta=\pm 1}\sum_{n=0}^{n_{max}}2\pi e|q_{i}|B\int_{-p_{i}^{(n,\eta)}}^{p_{i}^{(n,\eta)}}\rho_{i}\mathrm{d}p_{z}.
  \end{eqnarray}

\subsection{Coulomb interaction}
The size of strangelets with $A\ll 10^{7}$ is smaller than the electron Compton wavelength, thus electrons can not coexist with quarks in strangelets, in present calculations, the electron is ignored. Therefore, generally strangelets are not electric neutrality, this leads to a small Coulomb energy, in addition, the chemical potential of electron is treat as zero. Thus, combining with Eq.~(\ref{eqenergyE}), the energy density and energy per baryon of strangelets is given by
  \begin{eqnarray}
    E &=& \Omega+\sum_{i}\mu_{i}n_{i}+E_{C}+\frac{B^{2}}{2}  \label{eqenergy}\nonumber\\
     &=&\Omega_{0}+\sum_{i}\mu_{i}^{*}n_{i}+E_{C}+\frac{B^{2}}{2},   \\
    \frac{E}{n_{b}} &=& \frac{VE}{A}=\frac{4\pi R^{3}E}{3A}. \label{eqenergynb}
  \end{eqnarray}
  where $E_{C}$ is the Coulomb energy density, which is given by~\cite{Xia2014}
  \begin{equation}
    E_{C}=\frac{2}{15}\pi R^{2}\alpha (Q_{V}^{2}+5Q^{2}),
  \end{equation}
  where $\alpha\approx1/137$ is the fine structure constant and $Q_{V}$ is the volume term of the total electric charge density $Q$, i.e., $Q=\sum_{i}q_{i}n_{i}$ and $Q_{V}=\sum_{i}q_{i}n_{i,V}$. Here $n_{i,V}$ is given by
  \begin{eqnarray}\label{niv}
    n_{i,V}&=&\sum_{\eta=\pm}\sum_{n=0}^{n_{max}}2\pi e|q_{i}|B\int_{-p_{i}^{(n,\eta)}}^{p_{i}^{(n,\eta)}}\frac{g_{i}}{(2\pi)^{3}}\mathrm{d}p_{z}\nonumber\\
           &=&\sum_{\eta=\pm}\sum_{n=0}^{n_{max}}\frac{e|q_{i}|B g_{i}}{2\pi^{2}}p_{i}^{(n,\eta)}.
  \end{eqnarray}
  \par

 For strangelets, taking account into the contribution of Coulomb interaction, the parallel and
transverse pressures is given by
 \begin{eqnarray}\label{eqpressure}
 P_{\parallel}&=&-\Omega-V\frac{\partial \Omega_{0}}{\partial V}+P_{C}-\frac{B^{2}}{2}\nonumber\\
 &=&-\Omega_{0}-\frac{R}{3}\frac{\partial \Omega_{0}}{\partial R}-\sum_{i}n_{i}\mu_{I}+P_{C}-\frac{B^{2}}{2},
\end{eqnarray}
\begin{eqnarray}\label{eqpressureT}
 P_{\perp}=-\Omega-V\frac{\partial \Omega_{0}}{\partial V}+P_{C}+\frac{B^{2}}{2}-\mathcal{M}B.
\end{eqnarray}

In terms of the basic thermodynamic differential relation related to Coulomb interaction at zero temperature, i.e.,
\begin{equation}\label{pc}
  \mathrm{d}(VE_{C})=-P_{C}\mathrm{d}V+\sum_{i}\mu_{C,i}\mathrm{d}(n_{i}V),
\end{equation}
the pressure and chemical potential contributions of Coulomb interaction are given by~\cite{Xia2014}
\begin{eqnarray}\label{Coul}
  P_{C}&=&\frac{2}{9}\pi R^{2}\alpha\Big[\frac{3Q_{V}}{5\pi^{2}}\sum_{i}\frac{q_{i}g_{i}p_{F,i}^{2}}{n'_{i}}\nonumber\\
  &\times&\Big(
  n_{i}+\frac{R}{3}\frac{\partial n_{i}}{\partial R}-\frac{\partial n_{i}}{\partial m_{i}}\frac{\mathrm{d}n_{i}}{\mathrm{d}n_{b}}n_{b}\Big)
  +Q^{2}-Q_{V}^{2}\Big],\\
  \mu_{C,i}&=&\frac{4}{15}\pi R^{2}\alpha q_{i}\Big(5Q+\frac{g_{i}p_{F,i}^{2}Q_{V}}{2\pi^{2}n'_{i}}\Big)\nonumber\\
  &-&\frac{2\alpha R^{2}Q_{V}}{45\pi}\sum_{j}\frac{q_{j}g_{j}p_{F,j}^{2}}{n'_{j}}\frac{\partial n_{j}}{\partial m_{j}}\frac{\mathrm{d}m_{j}}{\mathrm{d}n_{b}}.
\end{eqnarray}
where $n'_{i}$ is defined by
\begin{equation}\label{np}
  n'_{i}\equiv\frac{\partial n_{i}}{\partial p_{F,i}}.
\end{equation}

Thus, the actual chemical potential of quark with flavor $i$ is given by
\begin{equation}\label{muimuI}
  \mu_{i}=\mu_{i}^{*}-\mu_{I}+\mu_{C,i}.
\end{equation}

\section{Density and/or temperature dependent particle masses}
\label{sec:mass}

 Originally, the quark mass in quark mass-density-dependent model is given by~\cite{Chakrabarty1989_PLB229-112}
 \begin{equation}\label{bag}
   m_{i}=m_{0,i}+\frac{B_{bag}}{3n_{b}},
 \end{equation}
 where $i$ goes over $u$, $d$ and $s$, $m_{0,i}$ is the current mass of quark with flavor $i$, $B_{bag}$ is the bag constant. Based on the in-medium chiral condensates and linear confinement, a cubic root scaling was derived~\cite{Peng1999_PRC61-015201}
 \begin{equation}\label{linear}
   m_{i}=m_{0,i}+\frac{D}{n_{b}^{1/3}},
 \end{equation}
 where $D$ is the confinement parameter, which are determined by stability arguments of Witten-Bodmer hypothesis. On this foundation, quark mass scaling containing the effects of linear confinement and one-gluon-exchange interaction was obtained~\cite{Chen2012}
 \begin{equation}\label{gluon}
   m_{i}=m_{0,i}+\frac{D}{n_{b}^{1/3}}-C n_{b}^{1/3},
 \end{equation}
 where $D$ is still the confinement parameter, and $C$ represents the effect of one-gluon-exchange interaction. Xia and Peng noted that the $m_{I}$ in Eq.~(\ref{gluon}) can be understood as a Laurent series of Fermi momentum. From this point of view, considering the perturbative effect, a new mass scaling was formulated as~\cite{Xia2014_PRD89-105027}
  \begin{equation}   \label{eqmc'}
    m_{i}=m_{0,i}+\frac{D}{n_{b}^{1/3}}+C^{'}n_{b}^{1/3},
  \end{equation}
  where $C^{'}$ represents the leading order perturbative interaction. \par
  To facilitate the comparison of the effects of one-gluon-exchange and perturbative effect on the properties of strangelets, we will unify formulas Eq.~(\ref{gluon}) and Eq.~(\ref{eqmc'}) in writing
  \begin{equation}   \label{eqmc}
    m_{i}=m_{0,i}+\frac{D}{n_{b}^{1/3}}+Cn_{b}^{1/3},
  \end{equation}
 where it corresponds to the perturbation and one-gluon-exchange effect respectively when $C>0$ and $C<0$.

\section{Properties of Magnetized Strangelets}\label{Properties}
At a given baryon number, $A$, with the quark number density $n_{i}$ given in Eq.~(\ref{number}), the baryon number conservation is given by
\begin{equation}\label{baryon}
  A=\frac{1}{3}\sum_{i}n_{i}V.
\end{equation}

In the absence of electron, the beta equilibrium condition of strangelets should be replaced by
\begin{equation}\label{chemical}
  \mu_{u}=\mu_{d}=\mu_{s}.
\end{equation}

In this section, we solve numerically the equation of mechanical equilibrium, $P_{||}=0$, with baryon number conservation conditions and the beta equilibrium, i.e., Eqs.~(\ref{baryon})-(\ref{chemical}). We use $Z$ to denote the electric charge(in units of $e$), then we have
\begin{equation}\label{eqcharge}
  Z=\sum_{i}n_{i}q_{i}V.
\end{equation}

In present calculations, we use the quark mass scaling given by Eq.~(\ref{eqmc}). Note that for a negative parameter $C$, the quark mass scaling Eq.~(\ref{eqmc}) is equivalent to the mass scaling considering one-gluon-exchange effect, i.e., Eq.~(\ref{gluon}). \par

\begin{figure}
\centering
\includegraphics[width=8cm]{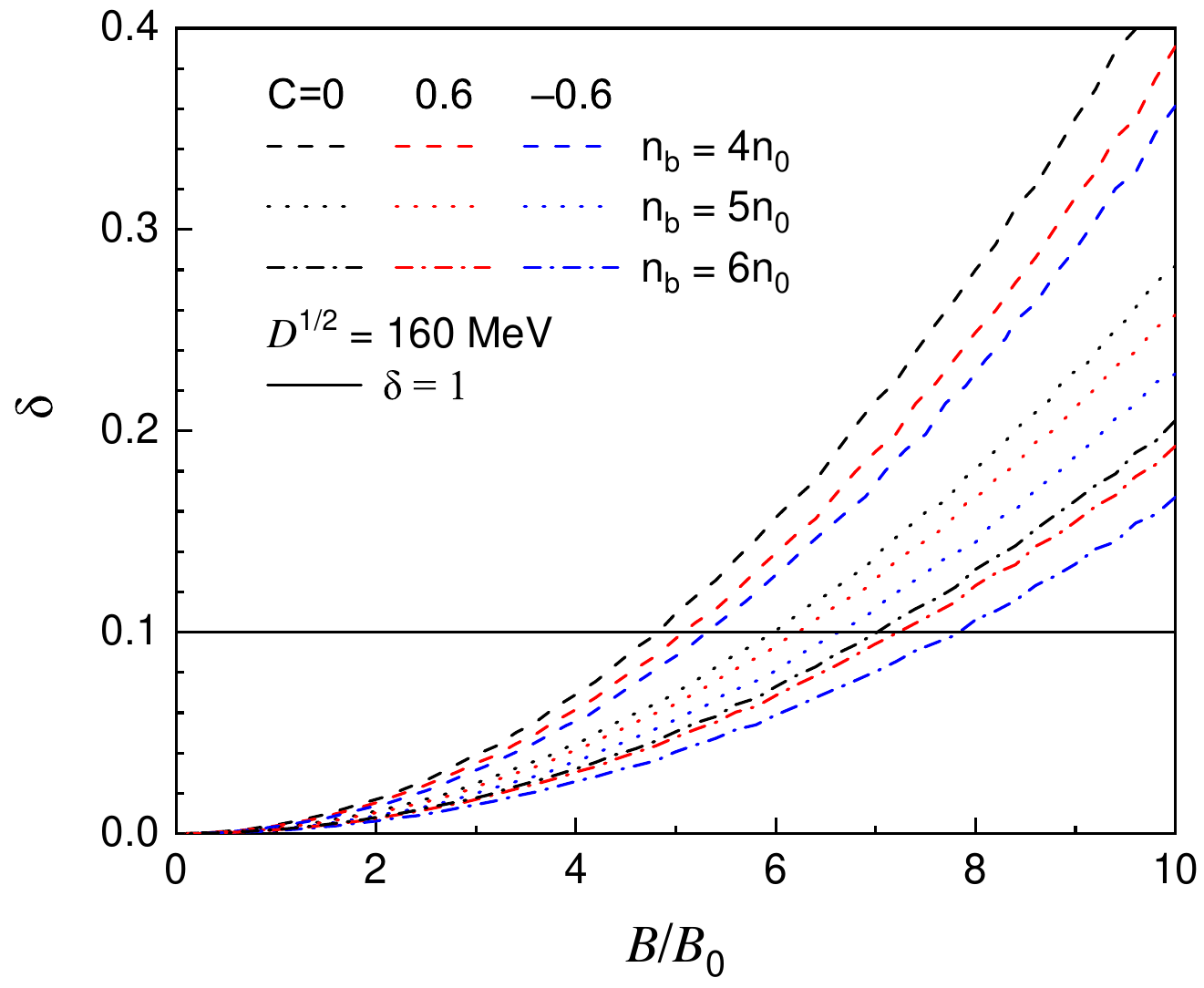}
\caption{The anisotropy of SQM as a function of magnetic field strength, where $B_{0}= 10^{17}\ \mathrm{G}$. }
\label{anisotropy}
\end{figure}

\begin{figure}
\centering
\includegraphics[width=8cm]{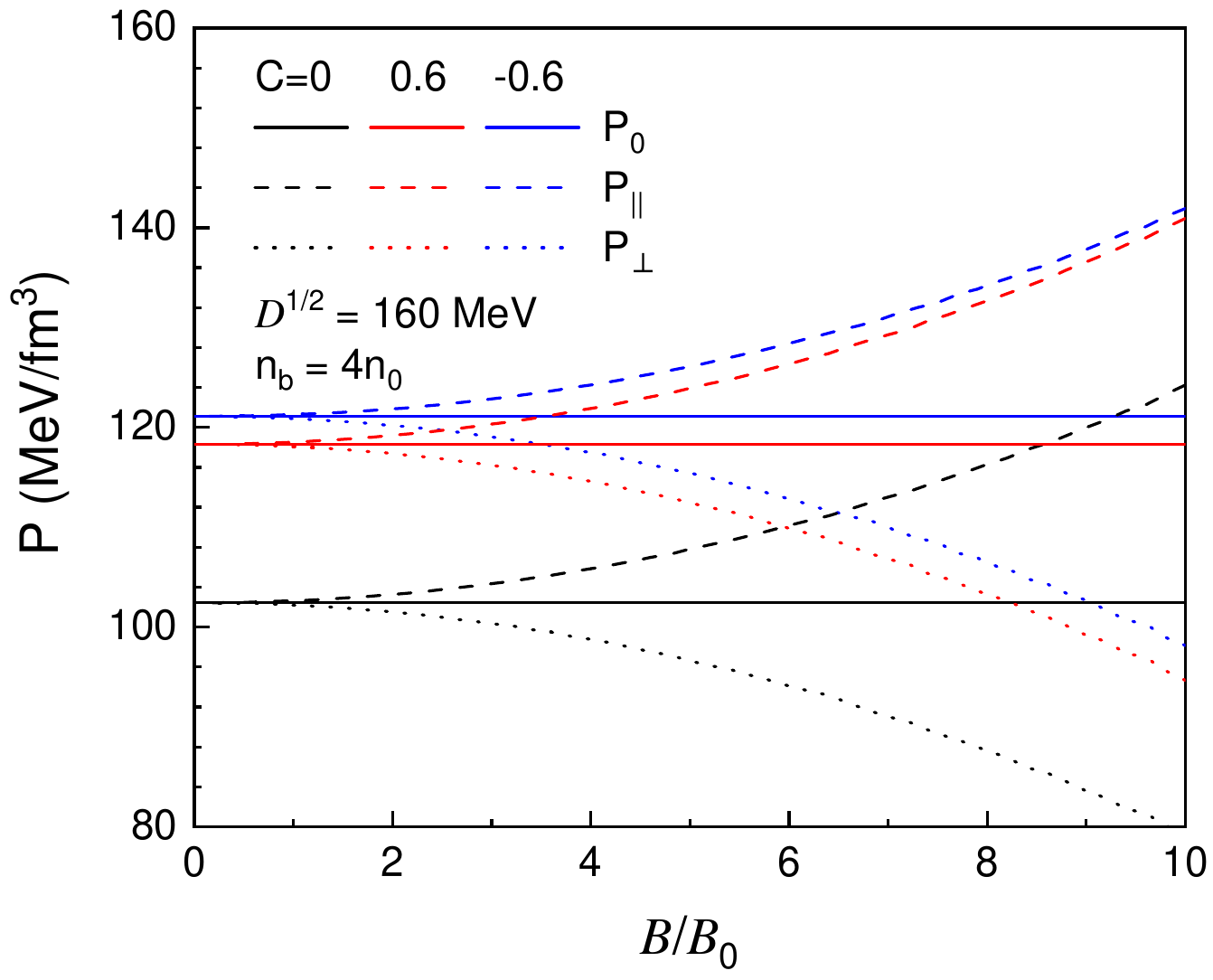}
\caption{The parallel and transverse pressures of SQM as a function of magnetic field strength at $n_{b}=4n_{0}$. The corresponding pressures $P_{0}$ at $B=0$ are included for comparison.}
\label{paratrans}
\end{figure}

In Fig.~\ref{anisotropy}, we present the anisotropy of SQM as a function of the field strength $B$ for fixed parameters $C$ and $D$. The dashed, dotted, dashed dotted and solid curves correspond, respectively, to the baryon number density $n_{b}=4n_{0}$, $5n_{0}$, $6n_{0}$ and $\delta=0.1$, where $n_{0}=0.165\ \mathrm{fm^{-3}}$ is the nuclear saturation density. It can be seen from the Fig.~\ref{anisotropy} that the anisotropy of SQM increases with the increase of magnetic field strength $B$, but the anisotropy is not significant for the pressure when the field strength $B\leq2\times 10^{17}\ \mathrm{G}$. In addition, the anisotropy of SQM decreases with increasing baryon number density. This is consistent with the conclusion of previous studies~\cite{Ferrer2010_PRC82-065802,Cui_NST2015.26.040503}. For different values of parameter $C$, we notice that the anisotropy of SQM take the maximum value when $C=0$. When $n_{b}$ is large, the anisotropy of SQM of $C>0$ is more clearly larger than that of $C<0$. There are some interesting points, which we explain in combination with Fig.~\ref{paratrans}. From Fig.~\ref{paratrans}, we can see that this is mainly because the pressure $P_{0}$ ($B=0$) of strange quark matter is minimum at $C=0$, centered at $C>0$, and maximum at $C<0$, respectively. At the same time, there is not much difference in the degree of deviation between $P_{\parallel}$ and $P_{\perp}$ from $P_{0}$.

\begin{figure}
\centering
\includegraphics[width=8cm]{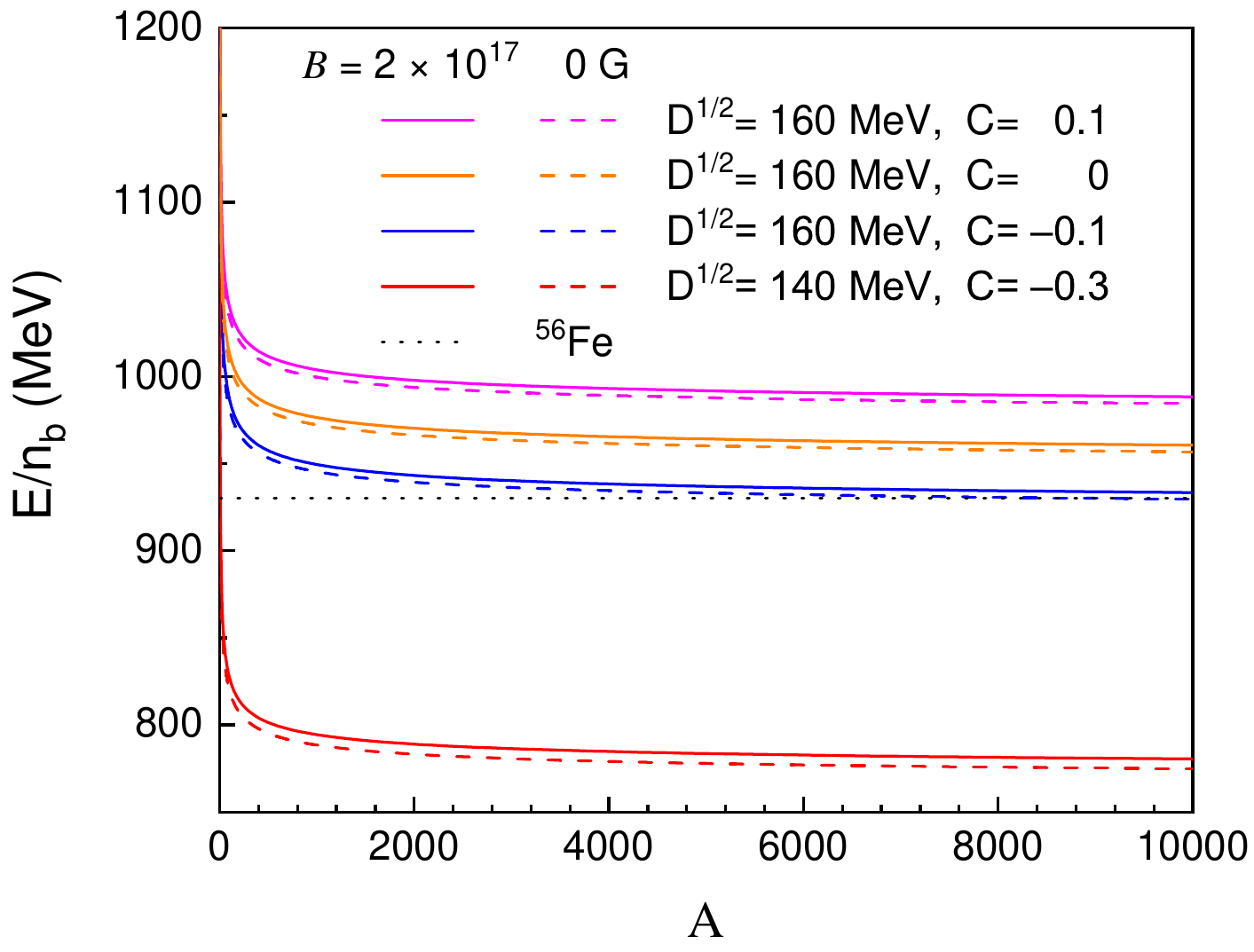}
\caption{The energy per baryon as a function of baryon number in both $B=0$ and $B=2\times 10^{17}\ \mathrm{G}$.}
\label{energy}
\end{figure}
\begin{figure}
\centering
\includegraphics[width=8cm]{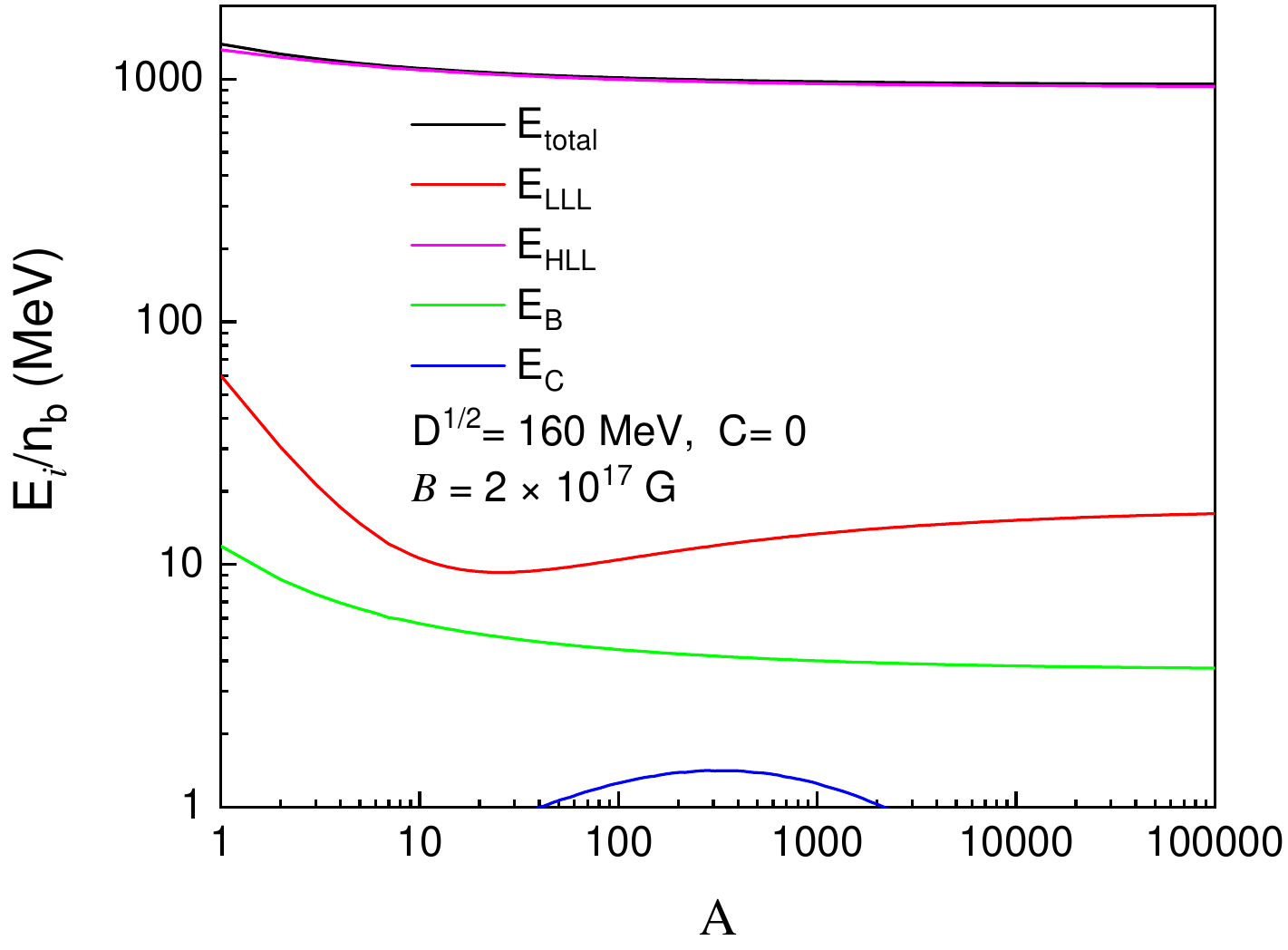}
\caption{The each component's energy per baryon $E_{i}/n_{b}$ as a function of baryon number in $B=2\times 10^{17}\ \mathrm{G}$.}
\label{energyi}
\end{figure}
Figures \ref{energy} shows the dependence of energy per baryon $E/n_{b}$ given by Eq.~(\ref{eqenergynb}) on the baryon number $A$ in both $B=2\times 10^{17}\ \mathrm{G}$ and $B=0$ for fixed parameters $C$ and $D$. The solid, dashed and dotted curves correspond, respectively, to $B=2\times 10^{17}$, $0$ $\mathrm{G}$ and $^{56}\mathrm{Fe}$. The magenta, orange, blue, and red curves correspond, respectively, to the parameter sets ($C$, $\sqrt{D}$ in $\mathrm{MeV}$): $(0.1,160)$, $(0,160)$, $(-0.1,160)$, and $(-0.3,140)$. It can be seen that the strangelets under $(-0.3,140)$ are absolutely stable at zero magnetic field strength, while the strangelets under $(-0.1,160)$ is metastable at $B=2\times 10^{17}\ \mathrm{G}$ when $A>8000$.
The energy per baryon decrease with increasing baryon number $A$.
This is consistent with Jensen and Madsen's results at $B=0$~\cite{JDM53prd4719}, and the value is slightly smaller.
This indicate that strangelets with large baryon numbers could be absolutely stable under strong magnetic field strength, which opens up the possibility of detecting strangelets from the particles emitted by unknown astrophysical sources that may have strong magnetic fields and high baryon numbers.
In addition, it is found that the first-order perturbation interaction($C>0$) increase energy per baryon, while the one-gluon-exchange interaction($C<0$) decreases energy per baryon, and the stronger one-gluon-exchange interaction, the lower energy per baryon.

\begin{figure}
\centering
\includegraphics[width=8cm]{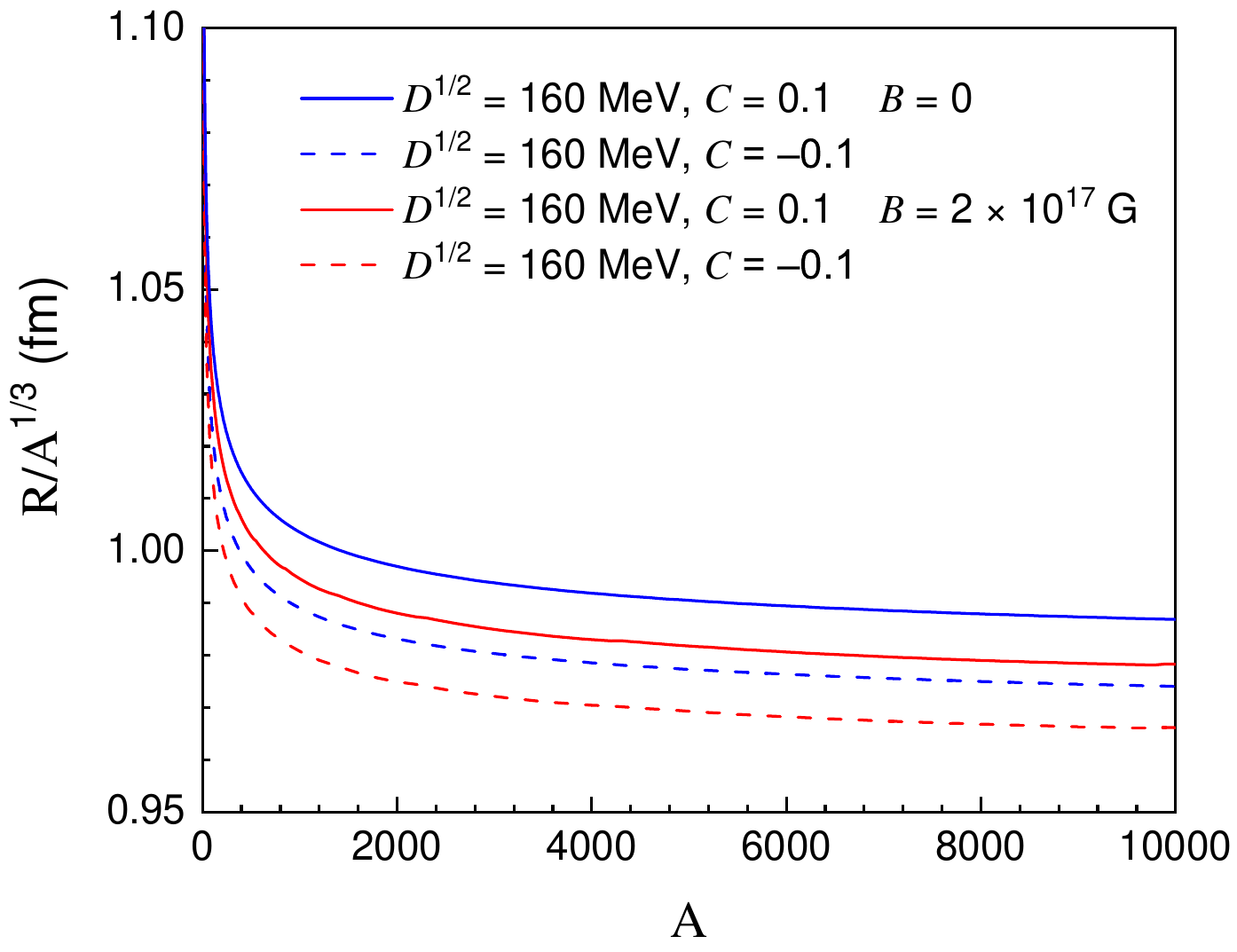}
\caption{The mechanically stable radius as a function of baryon number in both $B=0$ and $B=2\times 10^{17}\ \mathrm{G}$.}
\label{ratio}
\end{figure}
\begin{figure}
\centering
\includegraphics[width=8cm]{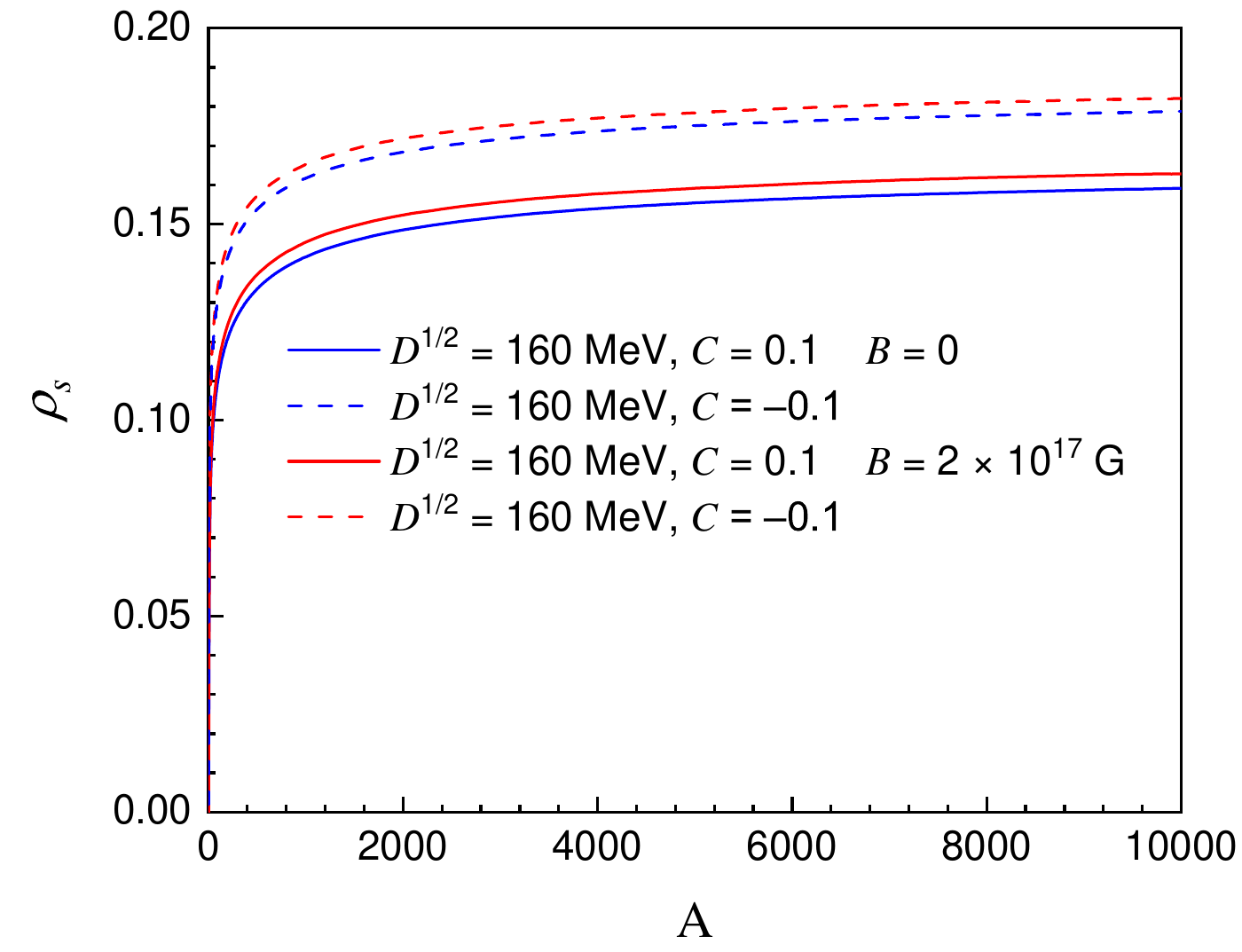}
\caption{The ratios of the strange quark number to tripling in baryon number $\rho_{s}$ as a function of baryon
number in both $B=0$ and $B=2\times 10^{17}\ \mathrm{G}$.}
\label{strange}
\end{figure}
The dependence of each component's energy per baryon on the baryon number is depicted in Fig.~\ref{energyi} in $B=2\times 10^{17}\ \mathrm{G}$ for fixed parameters $C$ and $D$. The black, magenta, red, green, and blue curves correspond, respectively, to the total energy per baryon $E_{total}/n_{b}$, lowest Landau level (LLL) energy per baryon $E_{LLL}/n_{b}$, higher Landau level (HLL) energy per baryon $E_{HLL}/n_{b}$, magnetic energy per baryon $E_{B}/n_{b}$ and Coulomb energy per baryon $E_{C}/n_{b}$, which the lowest Landau level energy $E_{LLL}$ and higher Landau level energy $E_{HLL}$ correspond to the $n=0$ and $n>0$ parts of the first two terms of Eq.~(\ref{eqenergy}), and the magnetic energy $E_{B}$ and Coulomb energy $E_{C}$ correspond to the third and fourth terms of Eq.~(\ref{eqenergy}). It can be seen that the higher Landau level energy $E_{HLL}$ accounts for the majority of the total energy and decreases as the number of baryons increases, however, the lowest Landau level energy $E_{LLL}$ decreases first and then increases, and tends to be stable. The magnetic energy $E_{B}$ decreases as the number of baryons increases. In addition, the Coulomb energy $E_{C}$ increases first and then decreases, and eventually tends to be zero. This is consistent with the electrically neutral of strange quark matter.\par
\begin{figure}
\centering
\includegraphics[width=8cm]{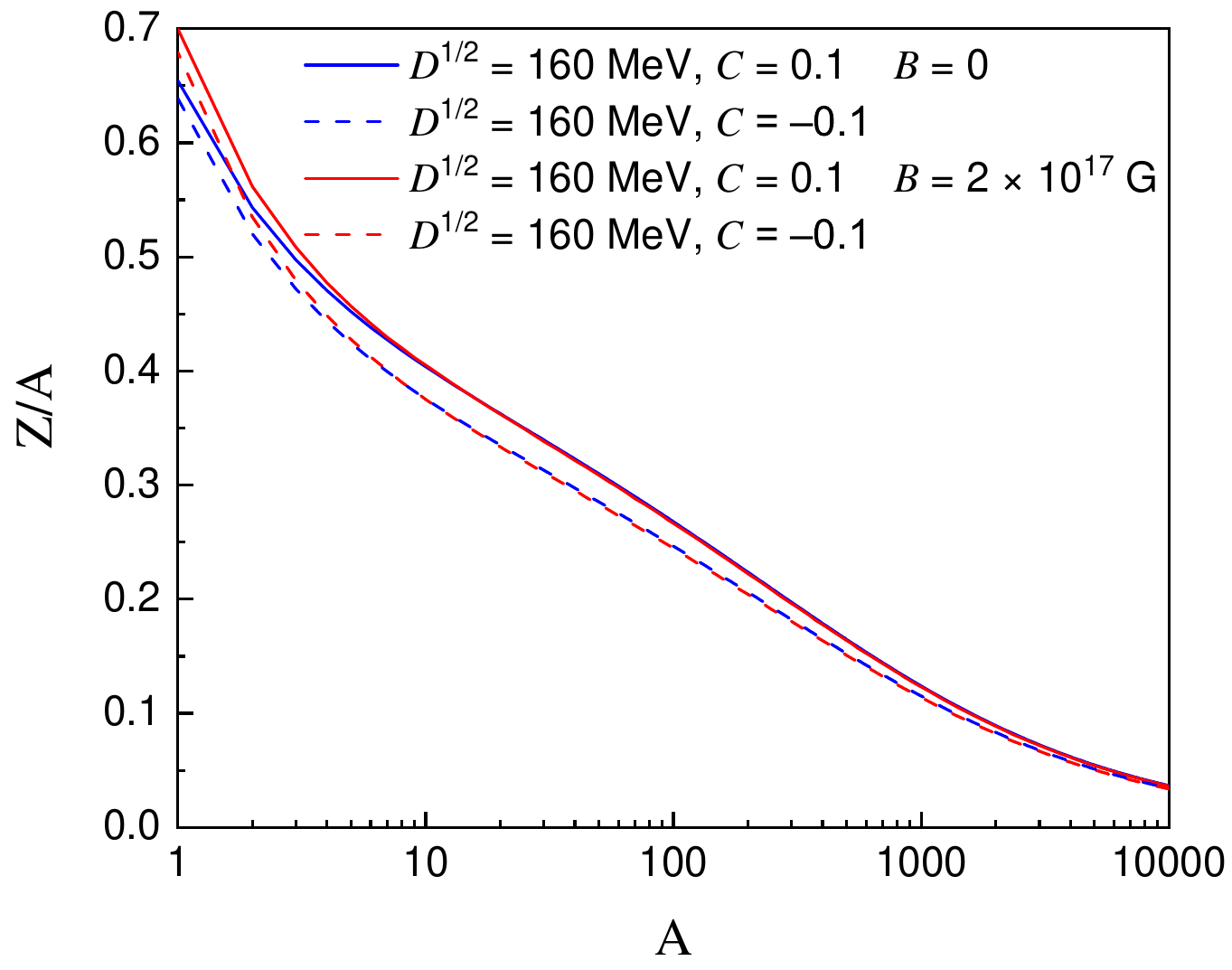}
\caption{The charge per baryon $Z/A$ as a function of baryon number.}
\label{charge}
\end{figure}
\begin{figure}
\centering
\includegraphics[width=8cm]{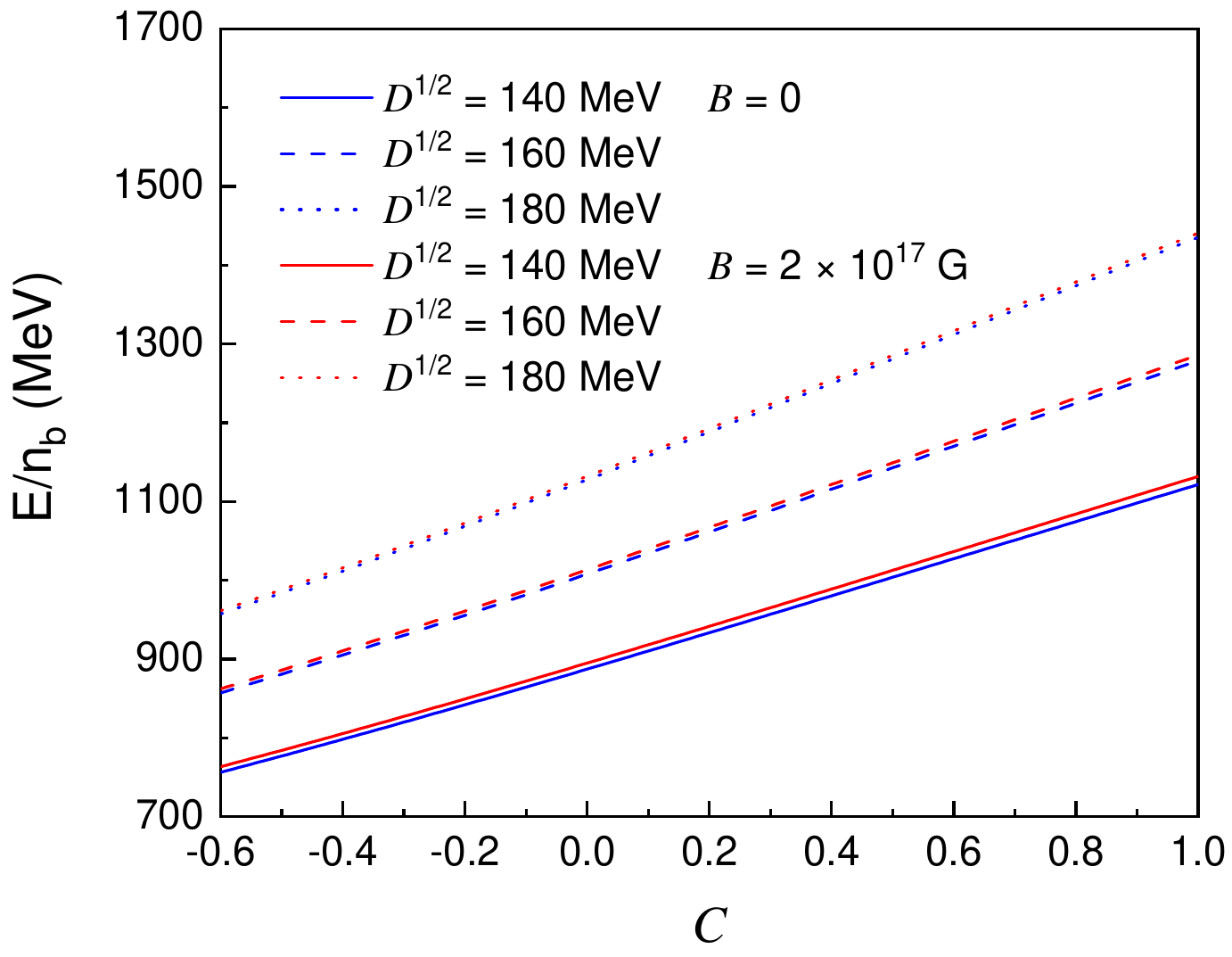}
\caption{The energy per baryon as a function of perturbative parameter with different confinement parameters.}
\label{cte}
\end{figure}
The dependence of mechanically stable radius of strangelets on the baryon number at $B=2\times 10^{17}\ \mathrm{G}$ and $B=0$ is shown in Fig.~\ref{ratio} by solving numerically the equation of mechanical equilibrium $P_{||}=0$ in Eq.~(\ref{eqpressure}), with baryon number conservation conditions and the beta equilibrium, i.e., Eqs.~(\ref{baryon})-(\ref{chemical}). The red and blue curves correspond, respectively, to the field strength $B=2\times 10^{17}\ \mathrm{G}$ and $B=0$. The solid and dashed curves correspond, respectively, to the parameter sets ($C$, $\sqrt{D}$ in $\mathrm{MeV}$): $(0.1,160)$ and $(-0.1,160)$. It can be seen that the ratio $R/A^{1/3}$ decreases with baryon number $A$ and tends to be a constant as $A$ approaches infinity, corresponding to an equation $R=r_{0}A^{1/3}$ with a constant $r_{0}$.
This is consistent with the conclusion of previous model studies such as NJL model~\cite{KO2005prd72}.\par

We use the ratio of the strange quark number to tripling of baryon number $\rho_s=N_s/3A$ to express the strangeness of strangelets, which is shown in Fig.~\ref{strange}. It is increasing with the baryon number, and tend to constant values at a large baryon number. Moreover, one-gluon-exchange interaction compared to perturbative interactions give a larger $\rho_{s}$. The strangeness of strangelets under a strong magnetic field is greater than that without a magnetic field.

\begin{figure}
\centering
\includegraphics[width=8cm]{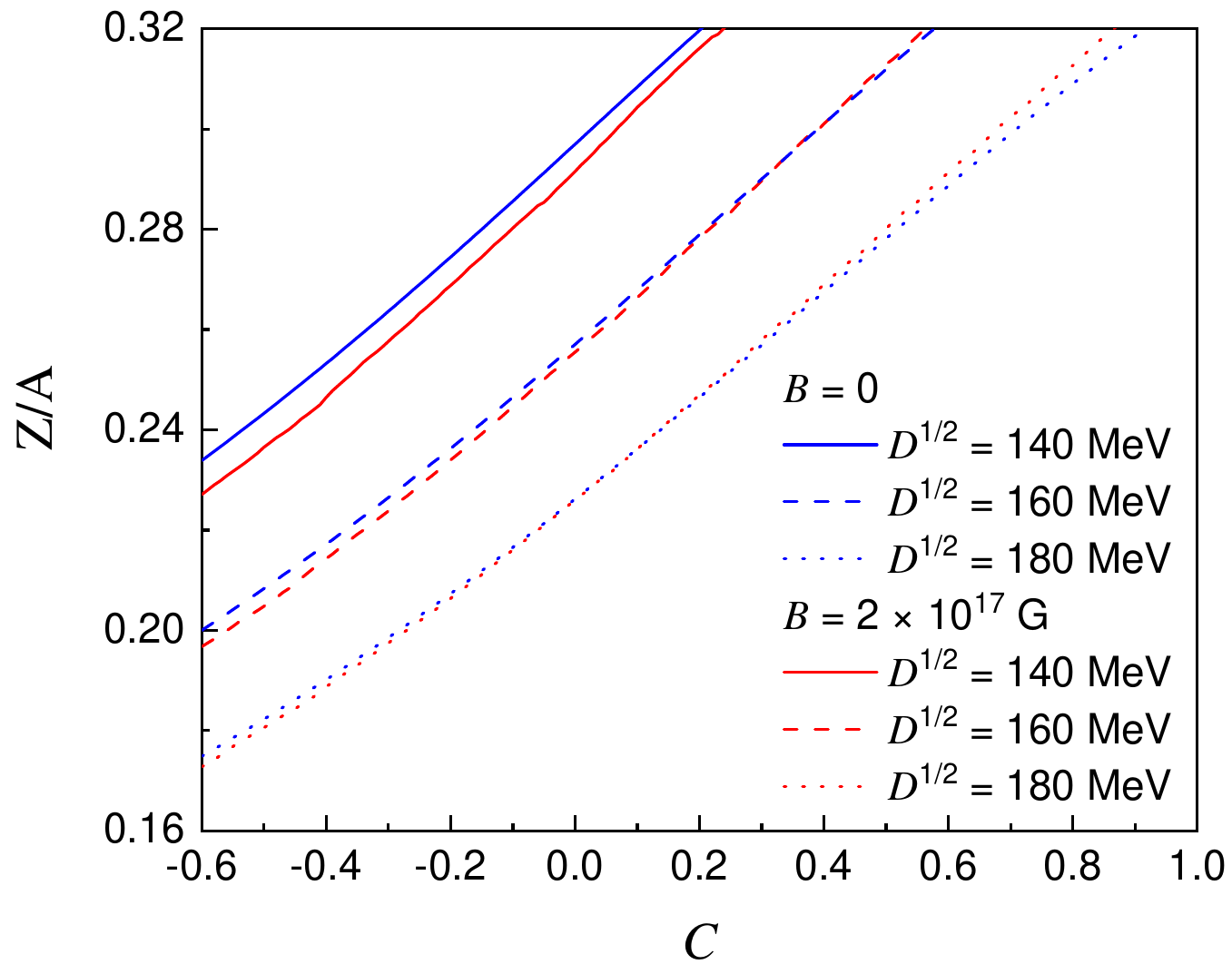}
\caption{The charge per baryon $Z/A$ as a function of perturbative parameter with different confinement parameters.}
\label{cpB}
\end{figure}
\begin{figure}
\centering
\includegraphics[width=8cm]{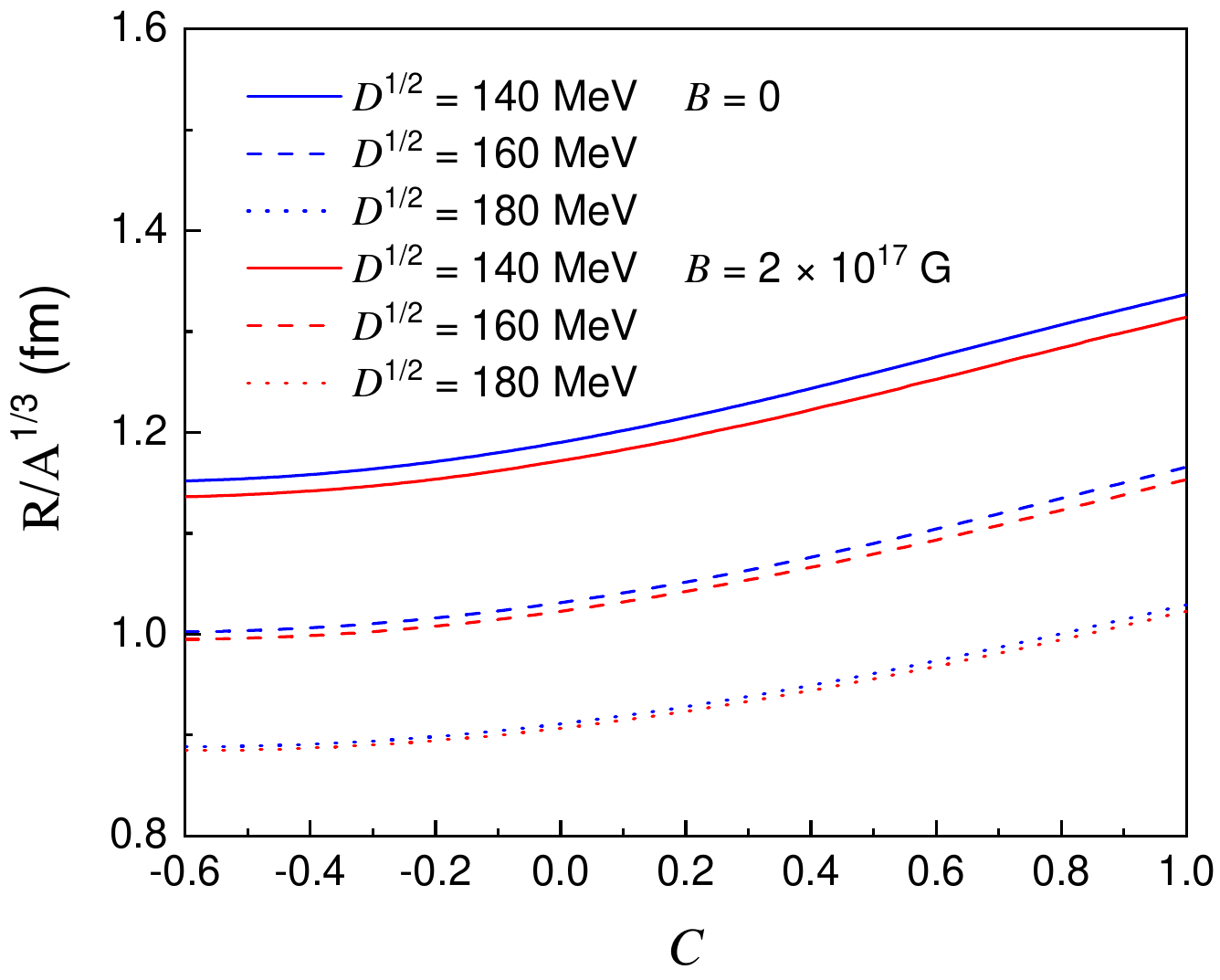}
\caption{The mechanically stable radius as a function of perturbative parameter with different confinement parameters.}
\label{ctr}
\end{figure}
\begin{figure}
\centering
\includegraphics[width=8cm]{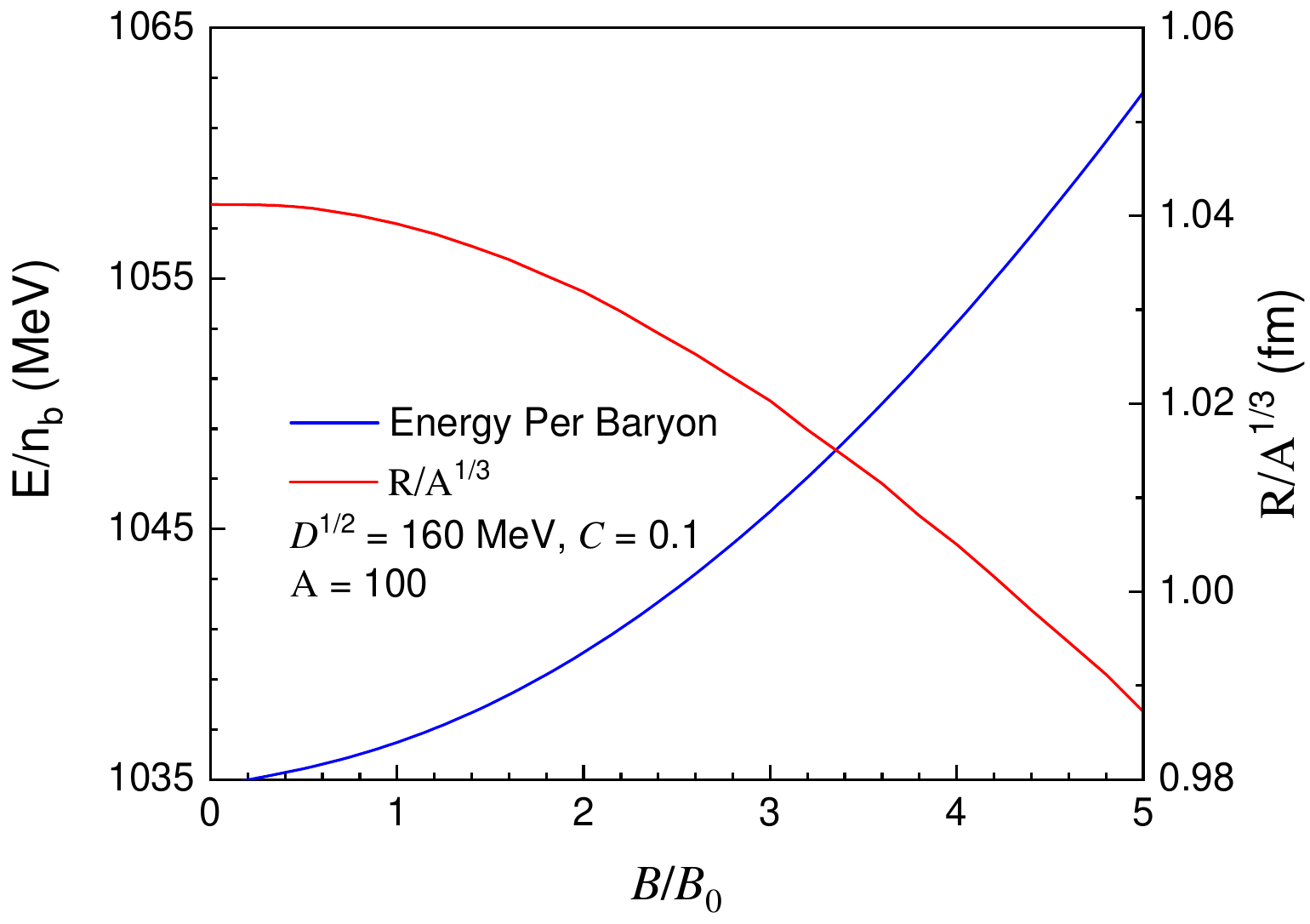}
\caption{The mechanically stable radius and energy per baryon as a function of magnetic field strength, where $B_{0}=10^{17}\mathrm{G}$.}
\label{B}
\end{figure}
\begin{figure}
\centering
\includegraphics[width=8cm]{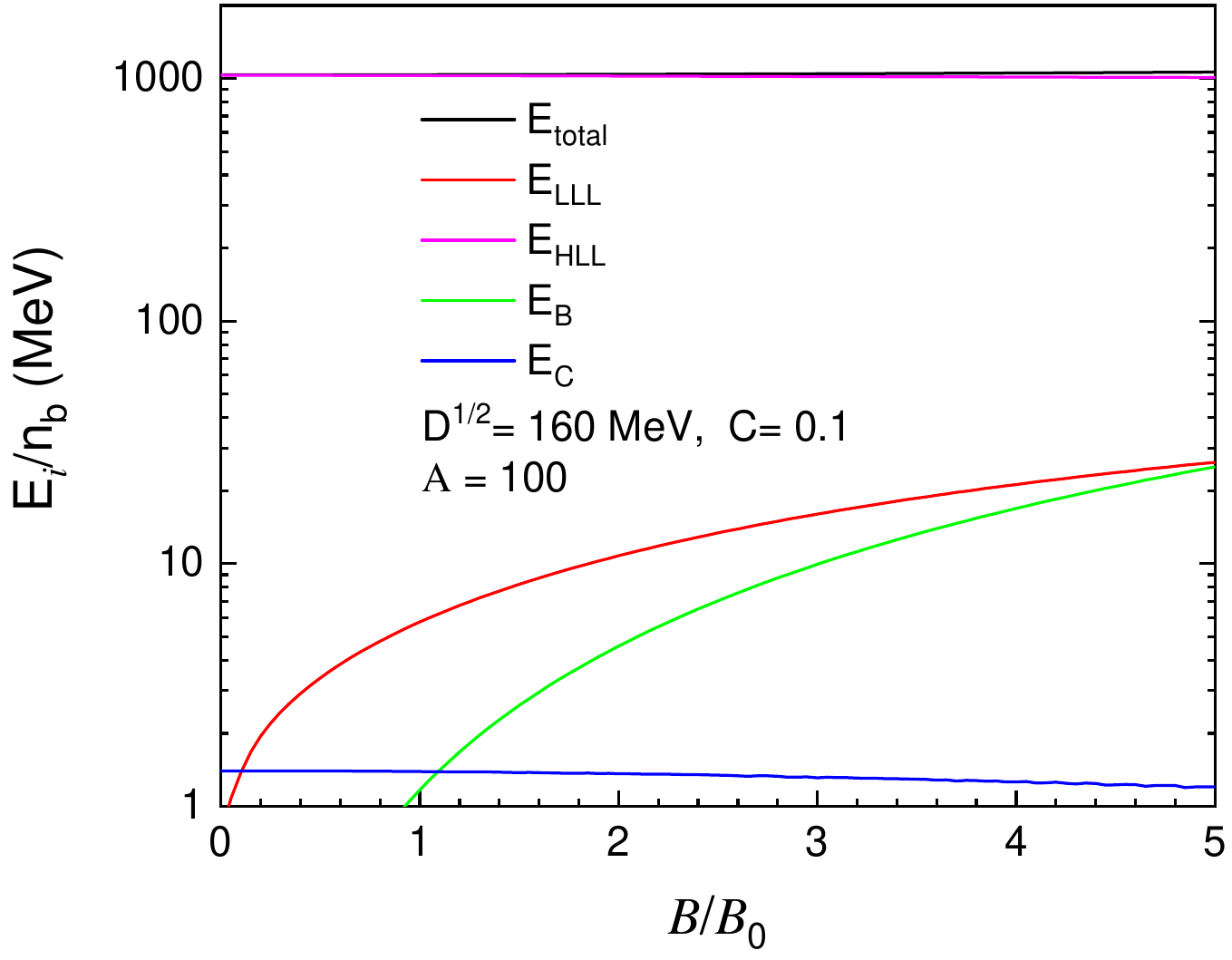}
\caption{The each component's energy per baryon $E_{i}/n_{b}$ as a function of magnetic field strength, where $B_{0}=10^{17}\mathrm{G}$.}
\label{BEi}
\end{figure}
The ratios of electric charge to the baryon number $Z/A$ as a function of baryon number $A$ is shown in Fig.~\ref{charge}. For a larger baryon number, generally, the strangelets and magnetized strangelets tend to be electrically neutral, while the finite size effect becomes weaker, i.e., they tend to be the SQM and MSQM. At a small baryon number, the radio $Z/A$ of magnetized strangelets is larger than the strangelets, however, with the increasing of $A$, they become close.

In Fig.~\ref{cte}, we note that the energy per baryon of strangelets and magnetized strangelets is a monotone increasing function of the perturbative parameter $C$ when $A$ is 100. Furthermore, we notice that the larger confinement interaction and strong magnetic field increases the energy per baryon of strangelets for $A=100$.

Next, the relationship between the charge per baryon of strangelets and magnetized strangelets and perturbative parameter $C$ is studied in Fig.~\ref{cpB}. It can be seen that the perturbative parameter $C$ increases the charge per baryon of strangelets in $B=0$ and $B=2\times 10^{17}\ \mathrm{G}$, and the rate of increase of magnetized strangelets is faster than strangelets.

The perturbative parameter $C$ dependence of the mechanically stable radii of strangelets and magnetized strangelets are presented in Fig.~\ref{ctr}.
The mechanically stable radius is increase with the increasing of the perturbative parameter $C$, both in $B=0$ and $B=2\times 10^{17}\ \mathrm{G}$, and a strong magnetic field will reduce the mechanically stable radii of the strangelets. This is consistent with the conclusion of previous model studies such as the MIT bag model~\cite{Ding2014_62-859}. In the model of the article, that also means the perturbative effects lead to larger mechanically stable radii of both strangelets and magnetized strangelets than the one-gluon-exchange effects. In addition, it is found that the stronger confinement interaction reduces the mechanically stable radii of strangelets and magnetized strangelets.

\begin{figure}
\centering
\includegraphics[width=8cm]{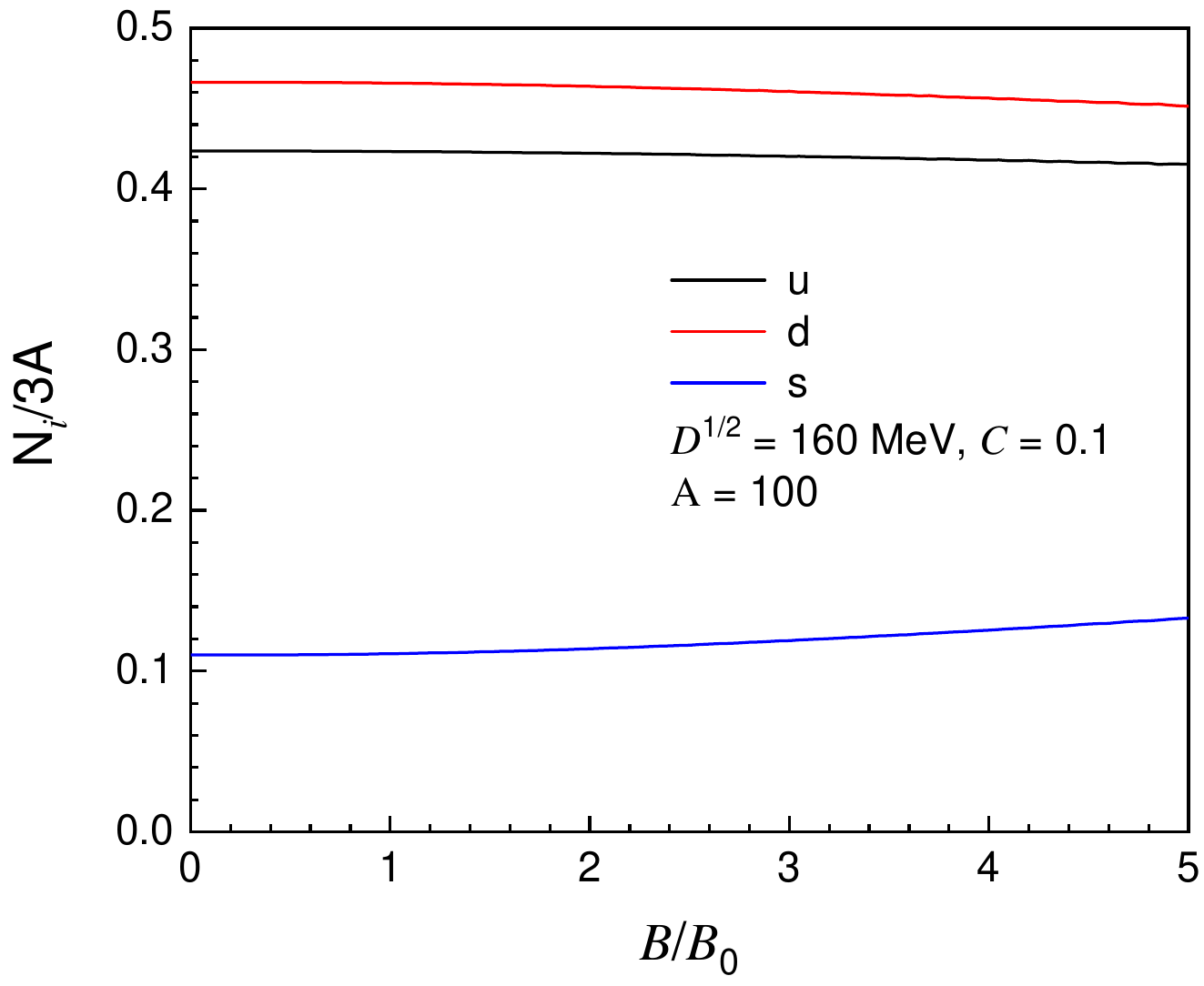}
\caption{The particle abundances $N_{i}/3A$ of quarks of the strangelets under fixed parameter set as a function of the magnetic field strength.}
\label{QB}
\end{figure}
The dependence of energy per baryon and the mechanically stable radius of magnetized strangelets on  the magnetic field strength are depicted in Fig.~\ref{B}. A parameter set $C=0.1$, $D^{1/2}=160 \mathrm{MeV}$ is choosen, where the energy per baryon increases with field strength, inversely, the mechanically stable radii decreases with field strength. This is consistent with the result of the energy per baryon and mechanically stable radii in Fig.~\ref{cte} and Fig.~\ref{ctr}.

The dependence of each component's energy per baryon on the magnetic field strength is depicted in Fig.~\ref{BEi} for fixed parameters $C$ and $D$. It is found that the higher Landau level energy $E_{HLL}$ accounts for the majority of the total energy and decreases as magnetic field strength increases, while the lowest Landau level energy $E_{LLL}$ is the opposite. This could be visualized from energy per baryon of strangelets at $n=0$ in Eq.~(\ref{eqenergy}) and the reduction of energy levels in Eq.~(\ref{Nmax}), respectively. The magnetic energy $E_{B}$ increases as the magnetic field strength increases. In addition, the Coulomb energy $E_{C}$ decreases with magnetic field strength, which is consistent with the phenomenon of the mechanical stability radius decreasing with magnetic field strength in Fig.~\ref{B}.

Figure~\ref{QB} shows the dependence of the particle abundances of quarks of the strangelets on the magnetic field strength.
The solid curves correspond to the parameter set $C=0.1$, $D^{1/2}=160 \mathrm{MeV}$.
The black, red and blue curves correspond, respectively, to the quark $u$, $d$ and $s$.
As the magnetic field strength increases, the particle abundance of $u$ and $d$ quarks decreases, while that of $s$ quarks increases because $u$ and $d$ quarks are converted to $s$ quarks under strong magnetic field strength.

\section{Summary}\label{Summ}
We have investigated the properties of magnetized strangelets by the baryon density-dependent quark mass model with a quark mass scaling contain confinement and perturbative effects. The contribution of Coulomb interaction has been treated in a thermodynamic self-consistent way, where the contribution of Coulomb interaction to the chemical potential and pressure is accounted for. Considering the anisotropy caused by a strong magnetic field and the anomalous magnetic moment of quarks, the thermodynamic quantities of the magnetized strangelets that satisfy thermodynamic consistency are obtained.\par
It is found that the dependences of energy and charge per baryon and mechanically stable radius of magnetized strangelets on the baryon number and perturbative parameter are similar to the strangelets in zero magnetic field.
The electric charge per baryon of strangelets and magnetized strangelets become close at a large baryon number, and both tend to electric neutrality.
However, the strong magnetic field leads to larger energy and charge per baryon, larger strangeness, and smaller mechanically stable radii of strangelets for fixed parameters, compared with the zero magnetic field one. For fixed confinement parameters, the energy and charge per baryon and mechanically stable radius increase as perturbative
parameter $C$ increases, and the rate of charge per baryon of magnetized strangelets is faster than strangelets.
In addition, the dependence of energy per baryon and mechanically stable radius on magnetic field strength are obtained, where the strong magnetic field lead to larger energy and smaller stable radius of strangelets with same baryon number.\par

\begin{appendix}
\section{Thermodynamics of quark matter in a uniform magnetic field}\label{sec:app}
In the presence of electromagnetic fields, the energy-momentum tensor can be decomposed into matter and field parts, i.e.,
\begin{equation}
T^{\mu\nu} = T^{\mu\nu}_{\rm matter} + T^{\mu\nu}_{\rm fields},
\end{equation}
where the field part is fixed by
\begin{eqnarray}
T^{00}_{\rm fields} &=& \frac{1}{2} (E^2+B^2),
\\
T^{0i}_{\rm fields} &=& T^{i0}_{\rm fields} = (\vec{E} \times \vec{B})_i,
\\
T^{ij}_{\rm fields} &=& \frac{1}{2} (E^2+B^2)\delta_{ij} - E_i E_j - B_i B_j.
\end{eqnarray}

To obtain the matter part of the energy-momentum tensor $T^{\mu\nu}_{\rm matter}\equiv {\cal T}^{\mu\nu}$ in the local rest frame of the system, we first write the Lagrangian density of the equivparticle model for quark matter, i.e.,
\begin{equation}
\mathcal{L} =  \sum_{i} \bar{\psi}_i \left[ i \gamma^\mu D_\mu  - m_i(n_\mathrm{b}) + \frac{1}{2} Q_i \sigma^{\mu\nu} F_{\mu\nu} \right]\psi_i,  \label{eq:Lgrg_all}
\end{equation}
where  $q_i$ is the charge of quark $i$, $Q_i$ the anomalous magnetic moment, $D_\mu=\partial_\mu + i q_i  A_\mu$, $m_i=m_{0i}+m_\mathrm{I}(n_\mathrm{b})$, and $\sigma^{\mu\nu} = i [ \gamma^\mu,\gamma^\nu ]/2$. The baryon number density is obtained with $n_\mathrm{b}=\sum_{i}\langle\bar\psi_i \gamma^0 \psi_i\rangle/3$. The equations of motion for $\psi_i$ can be determined by Euler-Lagrange equation, i.e.,
\begin{equation}
\frac{\partial {\cal L}}{\partial \bar\psi_i} - \partial_\mu \left( \frac{\partial {\cal L}}{\partial (\partial_\mu \bar\psi_i)} \right) = 0.
\end{equation}
This gives the Dirac equation for quarks, i.e.,
\begin{equation}
\left[i \gamma^\mu D_\mu - V_V \gamma^0 - m_i(n_\mathrm{b}) + \frac{1}{2} Q_i \sigma^{\mu\nu} F_{\mu\nu} \right]\psi_i = 0,
\label{eq:dirac}
\end{equation}
where the term $V_V$ arises from the density dependent quark masses, i.e.,
\begin{equation}
  V_V = \frac{1}{3}\frac{\mbox{d} m_\mathrm{I}}{\mbox{d} n_\mathrm{b}} \sum_{i} \langle\bar{\psi}_{i} \psi_{i}\rangle.
\end{equation}
The energy energy-momentum tensor is then obtained with
\begin{eqnarray}
{\cal T}^{\mu\nu}&=& \frac{1}{2}\sum_{i}\bar\psi_i \biggl[ i \left(\gamma^\mu D^\nu + \gamma^\nu D^\mu\right) \biggr] \psi_i  \nonumber \\
 &+& \frac{1}{2}\sum_{i}\bar\psi_i \biggl[Q_i \left(\sigma^{\mu\alpha} {F^\nu}_{\alpha} + \sigma^{\nu\alpha} {F^\mu}_{\alpha} \right) \biggr] \psi_i  \nonumber \\
 &-&  3 n_\mathrm{b}V_Vg^{\mu\nu}.
\end{eqnarray}
Note that we have adopted the relation ${\cal L}=3 n_\mathrm{b}V_V$ during derivation using the Minkowski space metric $g^{\mu\nu}={\rm diag}(1,-1,-1,-1)$, which is obtained multiplying Eq.~(\ref{eq:dirac}) by $\bar\psi_i$.

If the magnetic field is uniform and pointing to the $z$-direction, we can adopt the vector potential $A^\mu = (0,-By,0,0)$, which gives $F^{\mu\nu} = B (\delta^{\mu x}
\delta^{\nu y} - \delta^{\nu x} \delta^{\mu y})$ and consequently
\begin{equation}
\frac{1}{2} Q_i \sigma^{\mu\nu} F_{\mu\nu} = i Q_i B \gamma^x \gamma^y =
Q_i B \left( \begin{array}{cc} \sigma_3 & 0 \\ 0 & \sigma_3 \end{array} \right)
\equiv Q_i B {\cal S}_3.
\end{equation}
The Dirac equation for stationary states $\psi = e^{-i \varepsilon t} \Psi_i(\vec{x})$ can then obtained with
\begin{eqnarray}
\varepsilon \Psi_i &=& \left[-\vec{\alpha}\cdot \left(i \vec{\nabla} - q_i \vec{A}\right) + q_i A_0\right] \Psi_i \nonumber \\
 &+& \left[q_i A_0 + V_V + \beta m_i  - Q_i B \beta {\cal S}_3\right] \Psi_i. \label{eq:sdirac}
\end{eqnarray}
Here $\vec{\alpha} = \gamma^0 \vec{\gamma} = \left(\begin{array}{cc} 0            & \vec{\sigma}\\    \vec{\sigma} & 0 \\ \end{array}\right)$ and $\beta = \gamma^0 = \left(\begin{array}{cc} I  & 0 \\    0 & -I \\ \end{array}\right)$.
The diagonal components of ${\cal T}^{\mu\nu}$ in a constant magnetic field are then given by
\begin{eqnarray}
{\cal T}^{00} &=& \sum_i\bar\psi_i \left(i \gamma^0 D^0 \right) \psi_i -  3 n_\mathrm{b}V_V, \\
{\cal T}^{xx} &=& \sum_i\bar\psi_i \left(i \gamma^x D^x - Q_i B \sigma^{xy} \right) \psi_i +  3 n_\mathrm{b}V_V, \\
{\cal T}^{yy} &=& \sum_i\bar\psi_i \left(i \gamma^y D^y - Q_i B \sigma^{xy} \right) \psi_i +  3 n_\mathrm{b}V_V, \\
{\cal T}^{zz} &=& \sum_i\bar\psi_i \left(i \gamma^z D^z \right) \psi_i +  3 n_\mathrm{b}V_V.
\end{eqnarray}

For uniform matter comprised of $u$, $d$, and $s$ quarks, the electric field vanish. The energy energy-momentum tensor for a uniform magnetic field $B$ in parallel to the $z$-axis is determined by
\begin{equation}
  T^{\mu\nu}_{\rm fields} =\frac{B^2}{2} {\rm diag}(1,1,1,-1)
\end{equation}
The Dirac equation (\ref{eq:sdirac}) can be solved by assuming $\Psi_i(\vec{x}) = e^{i p_x x} e^{i p_z z} u^{(s)}_n(y)$ with
\begin{equation}
u^{(s)}_n(y) = \left(
\begin{array}{l}
c_1 \phi_\nu(y) \\
c_2 \phi_{\nu-1}(y) \\
c_3 \phi_{\nu}(y) \\
c_4 \phi_{\nu-1}(y) \\
\end{array}
\right)\ \ {\rm and} \ \ \nu = n + \frac{1}{2} - \frac{s}{2} \frac{q_i}{|q_i|},
\end{equation}
where $n=0,1,2,\cdots$ and the spin $s=\pm1$. The function $\phi_m$ is determined by
\begin{equation}
\phi_m(y) = N_m e^{-\xi^2/2} H_m(\xi) \ \ {\rm with} \ \  \xi = \sqrt{|q_i|B} \left( y + \frac{p_x}{q_i B} \right),
\end{equation}
where $m \geq 0$ is an integer, $H_m$ is a Hermite polynomial, and $N_m = (q_i B)^{1/4}(\sqrt{\pi} 2^m m!)^{-1/2}$
is a normalization constant which ensures $\int_{-\infty}^\infty dy \, \phi_n^2(y)~=~1$.
Inserting this to the Dirac equation (\ref{eq:sdirac}) gives
\begin{equation}
\left(
\begin{array}{cccc}
m - Q B & 0 & p_z & p_\nu \\
0 & m + Q B & p_\nu & -p_z \\
p_z & p_\nu & -m+Q B & 0 \\
p_\nu & -p_z & 0 & -m - Q B \\
\end{array}
\right)
\chi = E
\chi,
\end{equation}
where $\chi = (c_1 \; c_2 \; c_3 \; c_4)^{\rm T}$, $p_\nu = \sqrt{2 |q_i| B \nu}$, and $E = \varepsilon - q_i A_0 - V_V$. The eigenstates for particles are
\begin{equation}
\chi_s
= \frac{1}{\sqrt{2\lambda\alpha_s\beta_s}} \left(
\begin{array}{c}
s \alpha_s \beta_s \\
-p_z p_\nu \\
s \beta_s p_z \\
\alpha_s p_\nu
\end{array}
\right) ,
\label{spinorsol}
\end{equation}
where $\alpha_s \equiv E_s - Q_i B  + s \lambda$ and $\beta_s \equiv  \lambda + s m_i$. The corresponding eigenvalues are
\begin{equation}
E_s =  \sqrt{p_z^2 + (\lambda - s Q_i B)^2},
\end{equation}
with $\lambda \equiv \sqrt{m_i^2 + p_\nu^2}$.
The general positive energy states can be constructed by
\begin{equation}
    \psi_i(x)=\sum_{s=\pm 1}\frac{|q|B}{2\pi}\sum_n\int_{-\infty}^\infty \frac{dp_z}{2\pi}\frac{1}{\sqrt{2E_s}}b_s({\bf p})u^{(s)}({\bf p})e^{i\tilde{p}_\mu x^\mu},
\end{equation}
where $b_s({\bf k})$ obeys following relation
\begin{equation}
    \{b_r({\bf p}), b_s^\dagger ({\bf k})\}=(2\pi)\delta_{rs}\delta_{mn}\delta(p_z-k_z),
\end{equation}
${\bf p}=(n,k_z)$ with $n=0,1,2,\cdots$, $\tilde{p}=(E_s,p_x,0,p_z)$. For details, see Ref.~\cite{Strickland2012PRD86-125032}.

The energy density for infinite quark matter is then
\begin{eqnarray}
{\cal E} &=& \langle {\cal T}^{00} \rangle
 = \sum_i\frac{|q_i|B}{2 \pi} \sum_{s=\pm1}
\sum_n \int_{-\infty}^\infty \frac{d p_z}{2 \pi} \nonumber \\
& & \times E_s f_+(E_s,T,\mu_i^*), \label{eq:ener}
\end{eqnarray}
with the Fermi-Dirac distribution of particles fixed by
\begin{equation}
\langle b_s^{\dagger}({\bf p})b_s({\bf p})\rangle =f_+(E_s,T,\mu_i^*) = \frac{1}{e^{(E_s-\mu_i^*)/T} +1}.
\end{equation}
Note that we have adopted the relation $E_s = \varepsilon_s - q_i A_0 - V_V$ in the deriving Eq.~(\ref{eq:ener}). The number density can be determined by
\begin{equation}
n_i = \langle\bar\psi_i \gamma^0 \psi_i\rangle =  \frac{|q_i|B}{2 \pi} \sum_{s=\pm1} \sum_n \int_{-\infty}^\infty \frac{d p_z}{2 \pi}
 f_+(E_s,T,\mu_i^*),
\end{equation}
while the baryon number density is $n_\mathrm{b}=\sum_{i}n_i/3$.
According to Eq.~\eqref{eq:sdirac}, we have
\begin{eqnarray}
   & & (\varepsilon - q_i A_0 -V_V)\Psi_i^\dagger \Psi_i = \Psi_i^\dagger\left[-\vec{\alpha}\cdot \left(i \vec{\nabla} - q_i \vec{A}\right)\right] \Psi_i \nonumber \\
    &+& \Psi_i^\dagger\left[\beta m_i  - Q_i B \beta {\cal S}_3\right] \Psi_i.
\end{eqnarray}
Upon taking the derivative of the equation above with respect to mass, we obtain the following result:
\begin{equation}
    \Psi_i^\dagger \beta\Psi_i =\frac{\partial E}{\partial m_i}\Psi_i^\dagger\Psi_i,
\end{equation}
where $E=\varepsilon-q_iA_0-V_V$.
Note that $\bar{\psi}_i\psi_i=\psi_i^\dagger\gamma^0\psi_i$ and $\beta=\gamma^0$, we obtain
\begin{eqnarray}
\langle\bar{\psi}_i\psi_i\rangle &=& \frac{|q_i|B}{2\pi}\sum_{s=\pm}\sum_n\int^\infty_{-\infty}\frac{dp_z}{2\pi}\frac{\partial E_s}{\partial m_i}\langle b_s^{\dagger}({\bf p})b_s({\bf p})\rangle \nonumber \\
 &=& \frac{|q_i|B}{2\pi}\sum_{s=\pm}\sum_n\int^\infty_{-\infty}\frac{dp_z}{2\pi}\frac{\partial E_s}{\partial m_i}f_+(E_s,T,\mu^*_i). \nonumber \\
\end{eqnarray}
Thus, for vanishing temperatures with $T=0$, the expression of $V_V$ is given by
\begin{equation}\label{eq:vv_e}
    V_V=\frac{1}{3}\frac{\text{d}m_I}{\text{d}n_b}\sum_i\langle\bar{\psi}_i\psi_i\rangle=\frac{1}{3}\frac{\text{d}m_I}{\text{d}n_b}\sum_i\frac{\partial\mathcal{E}}{\partial m_i}.
\end{equation}

The longitudinal pressure along magnetic field lines is
\begin{eqnarray}
P_\parallel &=& \langle {\cal T}^{zz} \rangle \nonumber \\
&=& \sum_i
\frac{|q_i|B}{2 \pi} \sum_{s=\pm1} \sum_n \int_{-\infty}^\infty \frac{d p_z}{2 \pi}
\frac{p_z^2}{E_s} f_+(E_s,T,\mu_i^*) \nonumber \\
&+&  3 n_\mathrm{b}V_V.
\end{eqnarray}
The transverse pressure is fixed by
\begin{eqnarray}
P_\perp &=& \langle {\cal T}^{yy} \rangle = \langle {\cal T}^{xx} \rangle \nonumber \\
&=& \sum_i\frac{|q_i| B^2}{2 \pi^2} \sum_{s=\pm1} \sum_n \int_{-\infty}^\infty d p_z
\frac{1}{E_s} f_+(E_s,T,\mu_i^*) \nonumber \\
& & \times\Biggl[\frac{|q_i| \nu \bar{m}_i(\nu)}{\sqrt{m_i^2 + 2 \nu |q_i| B}} -   s Q_i \bar{m}_i(\nu) \Biggr] +  3 n_\mathrm{b}V_V, \nonumber \\
\end{eqnarray}
where $\bar{m}_i(\nu) \equiv \sqrt{m_i^2 + 2 \nu |q_i| B} - s Q_i B$.

For vanishing temperatures with $T=0$, the thermodynamic potential density is then fixed by $\Omega = {\cal E} - \mu_i n_i = - P_\parallel$ with $\mu_i = \mu_i^* + V_V$.
A relationship between $P_\parallel$ and $P_\perp$ obtained according to the magnetization $\mathcal{M} \equiv - \partial \Omega/\partial B = \partial P_\parallel/\partial B$, which gives $P_\perp = P_\parallel - \mathcal{M} B$.
These relations can be derived through direct integration, as detailed in Ref.~\cite{Strickland2012PRD86-125032}.

If we further include the field contributions to energy-momentum tensor, the energy density, longitudinal and transverse pressures at $T=0$ are fixed by
\begin{eqnarray}
{\cal E} &=& \sum_i\frac{|q_i|B}{2 \pi} \sum_{s=\pm1}
\sum_n \int_{-\infty}^\infty \frac{d p_z}{2 \pi} E_s \Theta(\mu_i^* - E_s)  \nonumber \\
&+& \frac{1}{2} B^2,
\end{eqnarray}
\begin{eqnarray}
P_\parallel = - \Omega -\frac{1}{2}B^2,
\end{eqnarray}
\begin{eqnarray}
P_\perp = - \Omega - \mathcal{M} B + \frac{1}{2}B^2 = P_\parallel + B^2 - \mathcal{M} B,
\end{eqnarray}
where
\begin{eqnarray}
\Omega &=& \sum_i\frac{|q_i|B}{2 \pi} \sum_{s=\pm1}
\sum_n \int_{-\infty}^\infty \frac{d p_z}{2 \pi} (E_s-\mu_i) \nonumber \\
& & \times\Theta(\mu_i^* - E_s),
\end{eqnarray}
\begin{eqnarray}
\mu_i = \mu_i^* + V_V,
\end{eqnarray}
\begin{eqnarray}
\mathcal{M} = - \frac{\partial \Omega}{\partial B}.
\end{eqnarray}
Due to the independence of chemical potential $\mu_i$ on mass $m_i$, combining with Eq.~\eqref{eq:vv_e}, we derive the relationship between $V_V$ and the thermodynamic potential density as follows
\begin{eqnarray}
    V_V=\frac{1}{3}\frac{\mathrm{d}m_I}{\mathrm{d}n_b}\sum_i\frac{\partial \Omega}{\partial m_i}.
\end{eqnarray}

For a finite system, the contributions from surface corrections and Coulomb interactions need to be considered, which introduce additional terms to the energy density and pressures.
\end{appendix}

\end{document}